\documentstyle[11pt,aaspp4]{article}

\begin{document}

\title{Observational Signatures of the First Luminous Objects}
\author{Siang Peng Oh\\
Princeton University Observatory, Princeton, NJ 08544; peng@astro.princeton.edu}

\begin{abstract}
The next generation of astronomical
instruments will be able to detect dense pockets of ionized gas created by the first luminous objects. The integrated free-free emission from ionized halos
creates a spectral distortion of the Cosmic Microwave Background,
greater than the distortion due to the reionized intergalactic
medium. Its detection is well within the capabilities of the planned DIMES satellite. Ionized halos may also
be imaged directly in free-free emission by the Square Kilometer
Array. Bright halos will be detected as discrete sources, while residual
unresolved sources can still be detected statistically from temperature fluctuations
in maps. Balmer line emission from ionized halos is detectable by the Next Generation
Space Telescope and can be used to obtain redshift information. Unlike
Ly$\alpha$, the H$\alpha$ line does not suffer from resonant scattering by
neutral hydrogen, and is still observable when the IGM is not fully
reionized. In addition, it is less susceptible to dust
extinction. Finally, the kinetic Sunyaev-Zeldovich effect may be used
to detect ionized gas at high redshift. Used in
concert, these observations will probe ionizing emissivity, gas
clumping, and star formation in the early universe.   
\end{abstract}

\section{Introduction}

Stimulated in part by the upcoming launches of the next generation of
astronomical instruments,
the epoch of reionization has become the subject of increasingly
intense theoretical effort in recent years. Reionization affects
observations of the Cosmic Microwave Background, both as a source of
new anisotropies, and in suppressing small-scale fluctuations from
the surface of last scatter (Haiman \& Knox, 1999, and references therein). In
addition, the Next Generation Space Telescope (NGST) offers the
prospect of directly detecting the first luminous objects. Nonetheless, the large
number of complex, non-linear physical processes
involved means that it is extremely difficult to make firm
theoretical predictions with any kind of confidence. While it appears
quite likely that NGST will be able to image Population III objects in
the infrared continuum (Haiman \& Loeb 1997, 1998), and detection of the
Gunn-Peterson trough may yield the redshift at which the IGM is not yet fully
reionized (Miralda-Escude \& Rees 1998), this information alone falls far short of being able to
fully constrain models of reionization. Key among the uncertainties is
the ionizing emissivity of collapsed objects, and the degree of gas
clumping. In this paper, we suggest the observation of diffuse gas and Population III objects in thermal bremstrahhlung and recombination line emission as a direct
probe of these quantities. Due to their $n_{e}^{2}$ dependence,
free-free and recombination line fluctuations trace the clumpiness
and ionization fraction of the gas. While the intensities of these
signals are directly proportional to one another, their
joint detection should prove invaluable, given the highly different
nature of the foreground contaminants and instrumental systematics in
each case. In addition, we examine the prospects for detecting ionized
gas at high redshift via the kinetic Sunyaev-Zeldovich effect. Below, we briefly summarize the observational case for each
signal.

{\bf Balmer line emission} Discussion in the literature on detecting line emission from high redshift galaxies has thus far
focussed predominantly on Ly$\alpha$ (Miralda-Escude \& Rees 1998,
Kunth et al 1998). However, detection in Ly$\alpha$ faces problems due
to attenuation by dust and resonant scattering by neutral
hydrogen. Indeed, in the dark ages before the universe was
completely reionized, it is unlikely that we will be able to detect
sources in Ly$\alpha$ emission, due to the large column density of
neutral hydrogen along the line of sight. By contrast, Balmer lines
face none of these problems, and provide a much cleaner probe of ionizing
emissivities. Thus, for instance, the intensity of
H$\alpha$ emission is intimately linked to star formation
rates. In this paper, we show that for equivalent integration times
with NGST,
moderate resolution spectroscopy (${\rm R \sim 1000}$) should yield H$\alpha$ line detection with the
same signal to noise as continuum IR imaging for $z<6.6$, and 10\%
signal to noise compared to IR imaging for $z>6.6$. Thus, if a high-redshift galaxy is
detected in imaging mode, H$\alpha$ line detection should be eminently
feasible. In addition, higher order Balmer lines such as H$\beta$
should also be detectable. The redshift information provided by Balmer line
emission should provide an invaluable complement to free-free and IR imaging.

{\bf Free-free emission} In addition to Compton scattering off hot gas via
the Sunyaev-Zeldovich effect, free-free emission from the ionized IGM
can also produce a spectral distortion of the
CMB (Barlett \& Stebbins 1991). This distortion increases
quadratically at low frequencies, and is poorly constrained by COBE observations near the peak of the CMB
spectrum ($\nu \sim 56$ GHz). One of the drivers behind the development of the DIMES
satellite (Kogut et al 1996; see http://ceylon.gsfc.nasa.gov/DIMES) is
the hope of detecting this distortion produced by the ionized IGM at
low frequencies ($\nu \sim 2$ GHz). The magnitude of the distortion is
directly related to the optical depth of ionized gas in the IGM, and
thus has the potential to constrain the epoch of
reionization.  However, the ionized IGM is not the only source of
free-free emission. Lyman limit systems in ionization
equilibrium with the external intergalactic ionizing background emit
thermal bremsstrahlung radiation (Loeb 1996), as do ionized halos with
an internal source of ionizing radiation (Haiman \& Loeb 1997). In
this paper, we argue that if the escape fraction of ionizing photons
from starburst galaxies is small (as is indicated by local
observations), then the integrated contribution of ionized halos to
the free-free background overwhelms that of the IGM, regardless of
the details of the reionization history of the universe. Thus, the
detection of a spectral distortion by DIMES is likely to probe the
integrated ionizing emissivity and clumping of gas in halos, rather
than the optical depth of the reionized IGM. 

A natural way to tell if an observed spectral distortion is produced by
ionised halos rather than an ionised IGM is to observe
fluctuations in the free-free background. If the spectral distortion
of the CMB is produced by the IGM, it should exhibit a fairly smooth
spatial signature across the sky, whereas if ionized halos predominate
then the free-free background should exhibit a great deal of small
scale power. Since Galactic foregrounds in free-free and synchrotron
emission damp rapidly at small scales, it should be possible to
directly image the predicted population of high redshift ionized
halos with a high resolution instrument. This is an application well suited for the proposed Square
Kilometer Array, or SKA (Braun et al 1998) which will operate in the
frequency range $0.03-20$ GHz with an angular resolution comparable
to that of HST ($\sim 0.1''$) and a sensitivity 100 times better than
the VLA, down to the $\sim 20$ nJy level. Such a quantum leap in
performance should reap immense scientific rewards. 

At present, the
appearance of the radio sky at nano Jansky levels is not known. We make a first stab at characterising
observable properties of a predicted population of free-free
emitters. In our Press-Schechter based model, we find that $\sim 10^{4}$ sources at $z>5$ should be
directly detectable as discrete sources in the $1^{\circ}$ SKA field of view. In
addition, the integrated background from sources too faint to be
directly identified creates a fluctuating signal, similar to the
Extragalactic Background Light (EBL)
in the optical and the Cosmic Infrared Background (CIB) in the IR,
which may be characterised by its
power spectrum and skewness. Finally, we investigate the clustering of
sources at high redshift and find that while it is fairly substantial,
the observed signal is still likely to be dominated by Poisson fluctuations.

In all numerical estimates, we assume a background
cosmology given by the 'concordance' values of Ostriker \& Steinhardt
(1995): $(\Omega_{m},\Omega_{\Lambda},\Omega_{b},h,\sigma_{8
h^{-1}},n)=(0.35,0.65,0.04,0.65,0.87,0.96)$. We use expressions from
Carroll, Press \& Turner (1992) and Eisenstein (1997) to compute
cosmological expressions such as the growth factor and luminosity distance.

\section{Modelling the Ionizing Sources}

\subsection{Estimating Halo Luminosities}

Since the free-free and H$\alpha$ emissivity are both proportional to
$\langle n_{e}^{2} \rangle$, observational prospects hinge crucially
on this quantity at high redshift. The signal is dominated by the small-scale
clumping of the gas at high redshift. One approach to estimating $\langle n_{e}^{2} \rangle$ is to model the properties of halos at high redshift. It is not known to what extent
the baryons can cool and condense in halos at high redshift. A minimal
assumption would be that gas within a halo traces the distribution of the dark
matter, generally overdense by $\delta \sim 180$ at collapse; this would
indicate $\langle n_{e}^{2} \rangle \propto (1+z)^{6}$. Observations
of the local universe indicate that the maximum luminosity
density of dynamically hot stellar systems varies as $M^{-4/3}$(Burstein et al 1997),
which implies that the gas is more centrally condensed in low mass
systems (which are prevalent at high redshift), presumably because dissipational effects are more
efficient. Thus, the above scaling with redshift is likely an
underestimate. If the gas settles to form
a disk it could be overdense by $\delta \sim 18 \pi^{2} \lambda^{-3}
\sim 10^{6}$ (Dalcanton, Spergel \& Summers 1997), where the spin
parameter $\lambda \sim
0.05$ typically (Barnes \& Efstathiou 1987). Note, however that $\langle
n_{e}^{2} \rangle$  is likely to be dominated by the local clumping of
the ISM in ionization bounded HII regions (in other words, the
clumping factor $\langle n_{e}^{2} \rangle/ \langle n_{e} \rangle^{2}$
of the gas is significantly greater than unity), in which case these limits
would have to be revised upwards.  

A simpler and likely more accurate approach is to assume that most
of the ionizing radiation produced is absorbed locally in
the halo, with only a small fraction escaping ($< 20\%$). This is
consistent with our observations of the nearby universe. For instance, the observation of 4 starburst galaxies with the Hopkins
Ultraviolet Telescope (HUT) by
Leitherer et al (1995) report an escape fraction of only $3\%$, by
comparing the observed Lyman continuum flux with that predicted from
theoretical spectral energy distributions. In our own galaxy, the
models of Dove, Shull \& Ferrara (1999) have an
escape fraction for ionizing photons of $6\%$ and $3\%$ (for coeval and
Gaussian star formation histories respectively) from OB associations
in the Milky Way disk, while Bland-Hawthorn \& Maloney (1999) find an escape fraction of $6\%$ is necessary for consistency with the observed line emission
from the Magellanic stream and high velocity clouds. The
increasing density of the gas in halos with collapse redshift indicates that the escape
fraction should decrease, strengthening our bound. The production rate of
recombination line photons $\dot{N}_{\rm recomb}$ is directly related to
the production rate of ionizing photons:
\begin{equation}
\dot{N}_{\rm recomb}= \alpha_{B} \langle n_{e}^{2} \rangle V \approx
(1-f_{esc})\dot{N}_{\rm ion}
\label{nesq}
\end{equation}
where $\dot{N}_{\rm ion}$ is the number of ionizing photons emitted per second,
and $\alpha_{B}$ is the case B recombination coefficient. 

In this manner the luminosities
of a source in H$\alpha$ and free-free emission scale directly with
the production rate of ionizing photons. For H$\alpha$, Hummer \& Storey (1987) find that
0.45 H$\alpha$ photons are emitted per Lyman continuum photon over a
wide range of nebulosity conditions.  We thus obtain directly
\begin{equation}
{\rm L(H \alpha)} = 1.4 \times 10^{41} \left(\frac{\dot{N}_{\rm ion}}{10^{53} {\rm
photons \, s^{-1}}} \right) {\rm ergs \, s^{-1}}.
\label{L_Halpha}
\end{equation}
For free-free emission, the volume
emissivity for a plasma is (Rybicki \& Lightman 1979):
\begin{equation}
\epsilon_{\nu} = 3.2 \times 10^{-39} n_{e}^{2} \left( \frac{T}{10^{4} K}
\right)^{-0.35} \ {\rm erg \,s^{-1} \,cm^{-3} \, Hz^{-1}}
\label{free_emissivity} 
\end{equation}
where we adopt the velocity averaged Gaunt factor $\bar{g}_{ff}=4.7$
and approximate its mild temperature dependence with a power
law. We can then use the estimate of $n_{e}^{2}V$
derived from equation (\ref{nesq}) to obtain:
\begin{equation}
L_{\nu}^{ff}=1.2 \times 10^{27} \left(\frac{\dot{N}_{\rm
ion}}{10^{53} {\rm photons \, s^{-1}}} \right) {\rm ergs \, s^{-1} \,
Hz^{-1}}
\label{L_free}
\end{equation}

We thus wish to estimate $\dot{N}_{\rm ion}(M)$, the production rate of ionizing
photons as a function of halo mass. We can either assume that this function
is independent of collapse redshift, or that it evolves with
redshift. In this paper, we adopt the models of Haiman \& Loeb (1998), which assume the former. In particular, we adopt their
starburst model, which is normalised to the observed metallicity $ 10^{-3} Z_{\odot} \le Z \le
 10^{-2} Z_{\odot}$ of 
the IGM at $z \sim 3$. In this model, a constant fraction of the gas
mass turns into stars, $1.7 \% \le f_{star} \le 17\%$ in a starburst which fades
after $\sim 10^{7}$ years. In this paper, we shall assume the upper
value $ Z \sim 10^{-2} Z_{\odot}$, corresponding to $f_{star} \sim
17\%$, in our discussion and plots. We show the results of the calculations
for the lower limit star formation efficiency $f_{star} \sim 1.7 \%$ in
figure (\ref{lower_SF_efficiency}) (discussed in section
(\ref{lower_SF_section})). During the starburst period, the
rate of production of ionizing photons is:
\begin{equation}
\dot{N}_{\rm ion}(M)=2 \times 10^{53} \, 
(f_{star}/0.17) \, (M/10^{9} M_{\odot}) \, {\rm photons} \,  {\rm s^{-1}}
\label{Ndot_scaling}
\end{equation}
Their mini-quasar model yields similar estimates, $\dot{N}_{\rm ion}(M)=4 \times 10^{53}
(M/10^{9} M_{\odot})$photons ${\rm s^{-1}}$, with the quasar fading after $\sim
10^{6}$ yrs. The
quasar model is normalised to the observed quasar luminosity function
at low redshift, assuming a constant ratio of black hole mass to halo
mass. We consider the adopted model to be a reasonably conservative
estimate for high redshift sources, as it seems quite possible that the
efficiency of ionizing photon production increases with redshift. The star formation rate is tied to
the dynamical timescale of the gas, SFR$\propto M_{g}/t_{dyn}$. Since high redshift sources are
denser, the free-fall time is shorter and $\dot{N}_{ion} \propto {\rm SFR}
\propto (1+z)^{1.5}$. Alternatively, one can use the empirical Schmidt law SFR$\propto \Sigma_{g}^{1.4}$
(Kennicutt 1998), where $\Sigma_{g}$ is the gas surface density, in
conjunction with an isothermal sphere model to deduce $\dot{N}_{ion} \propto {\rm SFR}
\propto (1+z)^{0.8}$ (all scalings are for an object of fixed collapse
mass). Intriguingly, Weedman et al (1998) find from
observations of galaxies in the HDF that high redshift starbursts have
intrinsic UV surface brightnesses typically four times brighter than
low redshift starbursts. Furthermore, the absence of metals to provide
cooling in the early universe would tend to lead to a top-heavy IMF
(Larson 1998), further increasing the efficiency of ionizing photon
production. Note also that in our adopted model the escape fraction
should fall with redshift, since $\dot{N}_{\rm ion}$ is independent of
redshift, whereas $\dot{N}_{\rm recomb} \propto \langle n_{e}^{2}
\rangle V \propto (1+z)^{3}$ (using $r_{vir} \propto (1+z)^{-1}$).  

We pause to note the possible caveat that supernovae could drive gas
out of the galaxy (Dekel \& Silk 1986), thereby drastically increasing
the escape fraction of ionizing photons and vitiating our
assumptions. The expulsion of gas would also shut off further star
formation. Although there is enough energy from supernovae to drive
such a wind, recent numerical simulations (Mac Low \& Ferrara
1999) indicate that mass
ejection is in fact only efficient in galaxies with $M_{g} < 10^{6}
M_{\odot}$ (such galaxies are excluded from our analysis), where the
small depth of the potential well means that gas is extremely easily
unbound. In higher mass galaxies, the supernovae produce a hole in the interstellar
medium through which the shock
heated gas is expelled, while the cold gas remains bound. We thus
ignore the effect of gas expulsion by supernovae. Note also that our
emissivities are calibrated 
to a quiescent, steady-state mode of star formation, as is observed in
the disk of many spiral galaxies today. There also exists the possibility
of massive starbursts, with star formation rates of up to 100 $M_{\odot} \,
yr^{-1}$, in which a large fraction of the gas mass is converted into
stars in the short dynamical time of the dense galactic center. This
is particularly likely to happen as halos merge, and would produce
significantly brighter objects, easily detectable with the proposed observations.    

When computing surface brightnesses, it is necessary to assume a
characteristic size for an ionized halo $r$. We can assume that the gas profile extends out to the virial radius ($r \sim
r_{vir}$) or that the gas is confined to a disk with spin parameter $\lambda$ ($r \sim \lambda r_{vir}$), where the virial radius $r_{vir}= v_{c}
/(10 \surd{2} H(z))= 2.1 {\rm Kpc} (M/10^{9} M_{\odot})^{1/3} (1+z/10)^{-1}$,
for an isothermal sphere. In subsequent calculations we shall assume $r
\sim \lambda r_{vir}$. The small corresponding angular size means that
in general we do not resolve the galaxy, which may thus be treated as
a point source. We quote angular size as $\theta \equiv {\rm
max(\theta_{telescope}},r/d_{A}(z))$, where $\theta_{\rm telescope}$
is the angular resolution of the telescope.  Note that this assumption only affects our
surface brightness calculations, since $L_{\rm H\alpha, \, free-free}
\propto \dot{N}_{\rm ion}$, independent of $r$.

\subsection{Statistics of Halos}
\label{halo_stat}

We employ a Press-Schechter based model of the abundance of ionizing
halos similar to that adopted by Haiman \& Loeb (1998). In our
simplified physical model, halos collapse, form
a starburst lasting $t_{o}=10^{7}$ years, and then recombine and no longer
contribute to the free-free background (FFB). We ignore
sources ionized by an external background radiation field, since the limiting column density before
these sources become self shielding corresponds to a undetectably
small flux at high redshift. We adopt a different expression for
the halo collapse from Haiman \& Loeb (1998); their expression
slightly underestimates the halo formation rate as it includes a negative
contribution from merging halos. The more accurate expression we use is
given by Sasaki (1994):
\begin{equation}
\frac{d\dot{N}^{form}}{dM}(M,z)=\frac{1}{D}\frac{dD}{dt}\frac{d n_{PS}}{dM}(M,z)\frac{\delta_{c}^{2}}{\sigma^{2}(M)D^{2}} 
\end{equation}
where $D(z)$ is the growth factor, and $\delta_{c}=1.7$ is the
threshold above which mass fluctuations collapse (note that at high
redshift, the difference between this expression and the one used by
Haiman \& Loeb (1998) is negligible, since mergers are unimportant). Here, $\frac{d\dot{N}^{form}}{dM}$
is the collapse rate of halos per mass interval per unit comoving volume, while
$\frac{d n_{PS}}{dM}$ is the standard Press-Schechter number density of collapsed halos per mass
interval (Press \& Schechter 1974, Bond et al 1991). Then, given an expression for the
expected flux from a halo of mass M at redshift z,  $S=S(M,z)$, we can calculate
the expected comoving number density of ionized halos in a given flux
interval as a function of redshift:
\begin{equation}
\frac{dN_{\rm halo}}{dS dV}(S,z)= \int_{t(z)}^{t(z)-t_{o}} dt
\frac{d \dot{N}^{form}}{dM} \frac{dM}{dS}
\end{equation}  
where $t_{o}$ is the duration of a starburst. 

Note, however, that star formation or quasar activity is only able to
proceed if gas is able to cool and condense in the halo. Thus, we set a cut-off mass for a halo to be ionized of
$M_{*}=10^{8} (1+z/10)^{-3/2} M_{\odot}$, which is the critical mass needed to
attain a virial temperature of $10^{4}$ K to excite atomic hydrogen
cooling. With this in mind, we can compute moments of the intensity distribution due to sources above a redshift $z_{\rm min}$ via: 
\begin{equation}
\langle S^{n}(>z_{min},S_{c}) \rangle= \int_{z_{min}}^{\infty} dz
\int_{S_{\rm min}(z)}^{S_{max}} dS
\frac{dN_{\rm halo}}{dSdV} \frac{dV}{dz d\Omega} S^{n}
\label{moments}
\end{equation}
where $S_{c}$ is a flux limit above which discrete source
identification and removal is possible. We shall use this expression entensively in ensuing
sections. The zeroth
moment of equation (\ref{moments}) can be used to compute the
number counts of sources above the flux limit $S_{c}$. In this case, $S_{max}
\rightarrow \infty$ and $S_{min}(z)={\rm max}(S_{c},S_{*}(z))$, where we
denote $S_{*}(z)$ as the flux from a halo of
minimum mass $M_{*}$ at redshift z. One can also use equation
(\ref{moments}) to compute the moments of the residual background
signal due to sources below the flux limit $S_{c}$, when all discrete
sources above the flux limit have been identified and removed. In this
case, we set $S_{max}=S_{c}$ and $S_{min}(z)=S_{*}(z)$.   

\subsection{Comparing Halo Emissivity with the IGM}

\begin{figure}
\epsscale{1.00}
\plotone{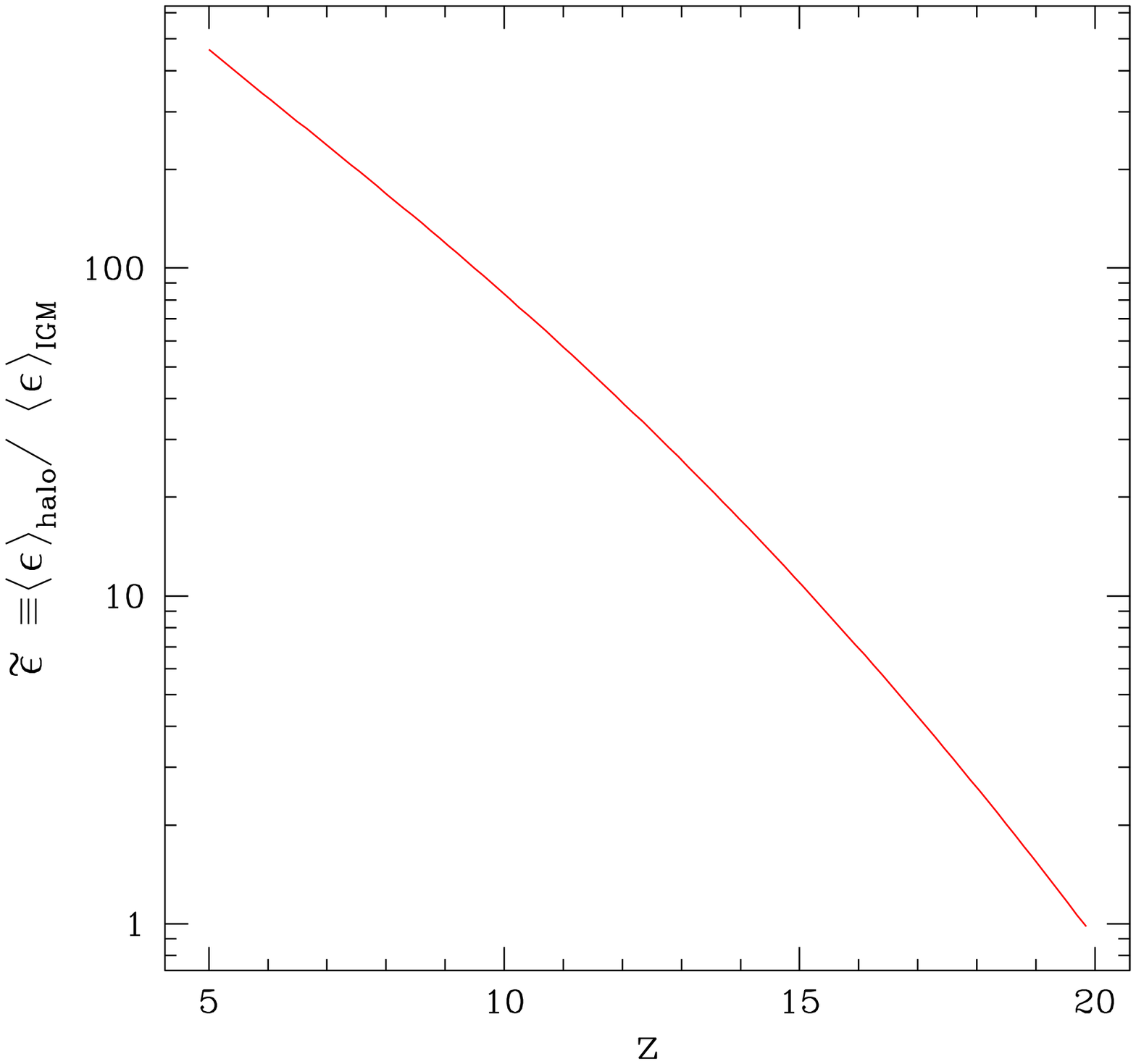}
\caption{The relative halo/IGM free-free emissivity $\tilde{\epsilon}\equiv \langle \epsilon \rangle_{\rm halo}/ \langle \epsilon \rangle_{\rm IGM}$ as a
function of redshift, assuming
that the entire uniform IGM is ionized. The contribution of halos to
the comoving free-free emissivity dominates that of the IGM at all
redshifts below 20.}
\label{clump}
\end{figure}

Throughout this paper we only compute the contribution of ionized
halos to free-free and recombination line emission, ignoring the emission from
filaments and the IGM. While it is obvious that small scale
fluctuations will be dominated by ionized halos, it is not immediately
obvious that the mean sky-averaged signal (as is probed by
spectral distortion measurements of the CMB) from ionized halos is greater than that
from other sources. This in fact is a direct consequence of our
assumption that the escape fraction of ionising photons from halos is
small. This implies that the degradation of ionizing photons into
recombination line photons take place largely in halos. The
fraction of ionizing photons produced which are degraded in halos as
opposed to the IGM (including ionized filaments and Lyman limit
systems) scales as $(1-\langle f_{esc}\rangle)/\langle f_{esc} \rangle $, where $f_{esc}$ is averaged over all
halos. This ratio is large due to the small assumed $f_{\rm
esc}$, which (as we've argued above) also is likely to decrease with
redshift. We can compare the volume averaged emissivities of the IGM
and ionized halos more quantitatively. The comoving free-free emissivity of the
ionized halos is given by:
\begin{equation}
\epsilon_{\nu}^{halos}(z)=\int_{M_{*}}^{\infty} dM \frac{dN}{dMdV}(M,z)
L_{\nu}(M,z) 
\end{equation}
where $L_{\nu}(M,z)$ is given by equations (\ref{L_free}) and
(\ref{Ndot_scaling}). The comoving free-free emissivity of the IGM may
be estimated directly from equation (\ref{free_emissivity}), assuming
the entire IGM is ionized. For the purpose of this
calculation, we have assumed a uniform IGM and ignored the
contribution of overdense structures without an internal source of
ionizing radiation. This is a good approximation as the low ambient
radiation field in the early universe means that these objects become
self-shielding at low column densities and are the last objects to
become reionized (Miralda-Escude, Haehnelt, \& Rees, 1998). In Figure
(\ref{clump}) we display the relative halo-IGM emissivity
$\tilde{\epsilon} \equiv \langle \epsilon \rangle_{\rm halo}/ \langle
\epsilon \rangle_{\rm IGM} = \langle n_{e}^{2} \rangle_{\rm halo}/
\langle n_{e}^{2} \rangle_{\rm IGM}$.  For $z<20$, the halo
emissivity dominates over the IGM emissivity (at high redshift, we
underestimate $\tilde{\epsilon}$ as the IGM is only partially
reionized. As computed by Haiman \& Loeb (1998), full reionization in
this model only occurs at $z \sim 13$). Note that this ratio
also applies directly to recombination line emission. We conclude that
the free-free and recombination line emission is dominated by discrete
sources rather than a diffuse background.

\section{Recombination line emission}

\subsection{H$\alpha$ emission}

The H$\alpha$ flux density from an ionized halo as detected with a R=1000
filter may be estimated as:
\begin{equation}
J_{\rm H \alpha}= \frac{L_{H \alpha}}{4 \pi d_{L}^{2}} \frac{1}{\Delta
\nu} \approx 625 \left( \frac{1+z}{10} \right)^{-1} \left
( \frac{R}{1000} \right) \left( \frac{\rm M} {10^{9} {\rm M_{\odot}}}
\right) {\rm nJy}
\label{J_Halpha}
\end{equation}
where we have estimated $L_{H \alpha}$ from equations (\ref{L_Halpha})
and (\ref{Ndot_scaling}). Note that the line width of H$\alpha$ is
simply given by Doppler broadening $\Delta \lambda / \lambda
\sim 10^{-4} (b/30 {\rm kms^{-1}})$ (unlike the Ly$\alpha$ line, which will
have broad damping wings), so it is still not resolved at
R=1000. Using a narrow bandpass enables us to reduce the amount of
background noise. Let us compare with the prospects for observability in
the infra-red, which corresponds to rest-frame UV emission. The UV continuum emission in the rest frame 1500--2800 \AA
(longward of the Lyman limit) for a Salpeter IMF is $L_{\nu} = 8 \times
10^{27} \left(\frac{{\rm SFR}}{\rm 1 M_{\odot} yr^{-1}}\right) = 8 \times 10^{27} \left(\frac{\rm \dot{N}_{ion}}  {\rm 10^{53} photons \, s^{-1}} \right) {\rm 
erg \, s^{-1} Hz^{-1}}$ (Madau 1998). Using equation (\ref{Ndot_scaling}), this
translates into an observed IR flux density:
\begin{equation}
J_{\rm IR}= \frac{L_{\nu}}{4 \pi d_{L}^{2}} (1+z) \approx 16 \left( \frac{1+z}{10} \right)^{-1} \left( \frac{\rm M}
{10^{9} {\rm M_{\odot}}} \right) {\rm nJy}
\label{J_UV}
\end{equation}
Hence, $J_{\rm H \alpha}/J_{\rm R}\approx 40 (R/1000)$. Note that the
spectral energy distribution/IMF of a luminous source can be probed by
its color, while the ratio of the H$\alpha$ luminosity to the rest
frame UV luminosity will constrain the escape fraction of ionizing
photons. 

Let us examine the competitiveness of H$\alpha$ with continuum IR imaging with
NGST. The signal to noise ratio of an observation with exposure time
$t$ is given by (Gillett \& Mountain
1998):
\begin{equation}
\frac{\rm S}{\rm N}=\frac{I_{s}t}{(I_{s}t +I_{bg}t+n\cdot I_{dc} \cdot t
+ n \cdot N_{r}^{2})^{1/2}}
\label{signal_to_noise}
\end{equation}
where $I_{s}=1.5 \cdot 10^{7} A \cdot \epsilon \cdot 1/R \cdot j_{\nu}
\cdot f$ electrons ${\rm s^{-1}}$ is the signal photocurrent ($A$ is
the area of the primary mirror in $m^{2}$, $\epsilon$
is the detector quantum efficiency ($\epsilon > 80\%$ for $\lambda < 5
\mu$m, $\epsilon > 50\%$ for $\lambda > 5
\mu$m), $j_{\nu}$ is the source strength in Jy, and f is the
fraction of source photons collected ($f \approx 0.7$ for $\lambda <
3.5 \mu$m and $f \approx 0.5$ for $\lambda >
3.5 \mu$m)),
$I_{bg}=1.5 \cdot 10^{7} A \cdot \epsilon \cdot 1/R \cdot \beta_{\nu}
\cdot \Omega$ electrons ${\rm s^{-1}}$ is the background photocurrent
($\beta_{\nu}$ is the background in Jy ${\rm sr}^{-1}$),
$n\approx 4$ is the number of pixels required to cover $\Omega \approx
0.01 \, {\rm arcsec^{2}}$, $I_{dc}$ is the detector
dark current, and $N_{r} < 15 \, e \, {\rm read^{-1}}$ (Stockman et al
1997) is the read noise per
pixel. We assume a point source subtends $\Omega \approx {\rm max}
(1,(\lambda/3.5 \mu m)) \, 0.01 \, {\rm arcsec^{2}}$. For $\lambda=0.5-5 \mu$m,
the projected dark current is $I_{dc} <0.02 {\rm e \, s^{-1}}$ for a
Si:As IBC array, while for
$\lambda > 5 \mu$m with an InSb array, $I_{dc} < 1{\rm e \, s^{-1}}$ (Stockman et al
1997). To estimate the sky background, we consider twice the sky background observed by
COBE (Hauser et al 1998). The sky background is measured at $1.25, 2.2,
3.5, 4.9, 12, 25, 60 \mu$m. We assume the full wavelength dependence
of the sky background in our calculations, employing a spline
interpolation between the observed points. Roughly speaking, the background is $< 10^{-5} \ {\rm Jy \
arcsec}^{-2}$ in the range $\lambda \sim 1-5 \mu$m, and rises to
$\sim 5 \times 10^{-4} \ {\rm Jy \
arcsec}^{-2}$ for $\lambda > 10 \, \mu$m. An additional
source of noise is thermal emission from the primary mirror, which we
approximate as a 75K blackbody (Barsony, 1999). This exceeds the
sky background at $\lambda \sim 10 \mu$m, and thus only affects
sources in H$\alpha$ emission with $z>15$. For IR continuum imaging ($R
= 3$) at $\lambda \sim 1-5 \mu$m, the sky
background is the dominant source of noise. Thus, we have ${\rm
S/N(IR)} \approx 0.5 (\Omega/{\rm 0.01 \, arcsec^{2}})^{-1/2} \left( \frac{1+z}{10} \right)^{-1} \left( \frac{\rm M}
{10^{9} {\rm M_{\odot}}} \right) t^{1/2}$. So, for instance, a 10$\sigma$ detection of our
canonical
$10^{9} M_{\odot}$ source at z=9 would take 400s. For spectroscopic
applications with $R>1000$, the detector noise is the dominant source
of noise. The increase in dark current at $\lambda > 5 \mu$m
corresponds to reduced a signal to noise ratio for sources with
$z>6.6$, leading to ${\rm S/N(H \alpha; z<6.6, z>6.6)} \approx (0.7,0.06) \, (\Omega/{\rm 0.01 \, arcsec^{2}})^{-1/2} \left( \frac{1+z}{10} \right)^{-1} \left( \frac{\rm M}
{10^{9} {\rm M_{\odot}}} \right)
t^{1/2}  $. A 10$\sigma$ detection of our
canonical
$10^{9} M_{\odot}$ source at z=9 in H$\alpha$ would take
$3 \times 10^{4}$s. Note that since the flux density $j_{\nu}^{H \alpha} \propto R$, the signal
to noise is independent of the spectral resolution $R$ in the dark
current limited regime $R>1000$. Thus, even higher resolution
spectroscopy, up to the point when the $H \alpha$ line is resolved ($R
\sim 10^{4}$),
should be possible. We derive
the useful rule of thumb that:
\begin{equation}
\frac{\frac{\rm S}{\rm N}({\rm H} \alpha)}{\frac{\rm S}{\rm N}({\rm
IR})}(z<6.6,z>6.6) = (1.4, 0.12) \left( \frac{t_{{\rm H} \alpha}}{t_{\rm IR}}
\right)^{1/2}
\label{halpha_vs_uv}
\end{equation}
Thus, for sources which sit at $z<$6.6, a source may be detected in IR
imaging and H$\alpha$ spectroscopy in equal
amounts of time. For sources at z$>$6.6, when the detector dark current for H$\alpha$ detection is signficantly larger, detection in H$\alpha$ requires
an integration time 100 times longer than in IR imaging to achieve the same
signal to noise ratio. This ratio will be reduced if advances in
technology permit smaller dark currents at $\lambda > 5 \mu$m. 

\begin{figure}
\epsscale{1.00}
\plotone{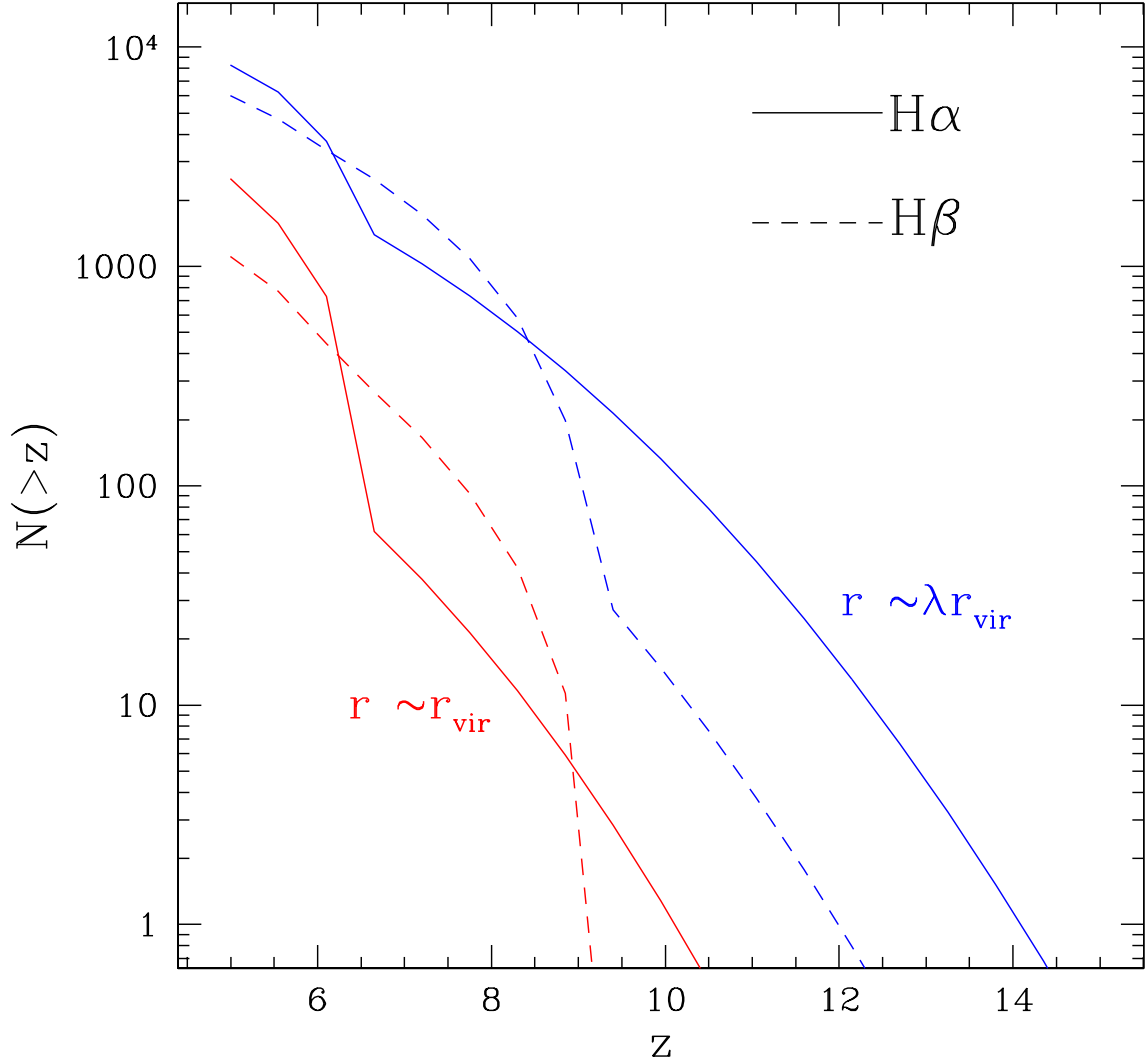}
\caption{Number of sources in $4' \times 4'$ NGST field of field with
a redshift greater than z, detectable with H$\alpha$, H$\beta$ spectroscopy (solid and dashed lines respectively) as a 10$\sigma$ fluctuation with
$t=10^{4}$s, and R=1000, for emission from a disc ($r \sim \lambda
r_{vir}$, essentially a point source), and an extended source ($r \sim
r_{vir}$). The dip at z=7,9 for H$\alpha$, H$\beta$ is due to increased detector noise at longer wavelengths $\lambda > 5 \, \mu$m. Note therefore the existence of an interval where H$\beta$ is potentially easier to detect than H$\alpha$. }
\label{Halpha_fig}
\end{figure}

Note that the detection rate
depends on the angular extent of the sources, even in the dark
current limited regime. An extended source covers more pixels,
resulting in a higher dark current (in very extended sources, it may
be best to focus on the central pixels, which contain most of the
light). For this reason, we quote the relative signal to noise in
equation (\ref{halpha_vs_uv}), which is independent of the angular
size. In figure (\ref{Halpha_fig}), we
show the number of sources detectable with a S/N of 10 with $R>1000$ spectroscopy in $10^{4}$s ($\sim 3$ hours) of integration time, all computed using the
full expression (\ref{signal_to_noise}), for both disk-like ($r\sim
\lambda r$) and extended ($r \sim r_{vir}$) objects. At least several
thousand objects with $z>5$ may potentially be detected in the NGST
$4' \times 4'$ field of view. The actual detection rate, of course,
depends on the targeting strategy for spectroscopy (see discussion
below). Note that if the star formation efficiency is lower, then the
detection rate falls. In figure (\ref{lower_SF_efficiency}), we show
the detection rate for $f_{star} \sim 1.7\%$, rather than $f_{star}
\sim 17\%$. For a star formation efficiency lower by an order of
magnitude, an integration time 100 times longer is required to obtain
the same number counts. 

It should be possible to observe higher order Balmer lines as
well. For case B recombination with $n_{e}=100 \, {\rm cm^{-3}}$ (the
density dependence is fairly weak), $f_{\rm H \beta}/f_{\rm H
\alpha}=0.35$ (Osterbrock 1989). Accounting also for the different $\delta \nu= \nu/R$ for H$\alpha$ and H$\beta$, we find
that $S/N({\rm H \beta})/S/N({\rm H \alpha})=0.25$ for a given source,
which implies an integration time 16 times longer is required to
observe a source in H$\beta$ with the same signal to noise ratio as in
H$\alpha$ (note that we assume that attenuation effects are still unimportant for H$\beta$, which is of shorter wavelength than H$\alpha$ and thus more suspectible to dust extinction or stellar absorption). However, in the redshift range $6.6<z<9.5$, the H$\beta$ line
falls in the regime $\lambda < 5 \mu$m with low dark current, while
the H$\alpha$ line falls in the $\lambda > 5 \mu$m regime with high
dark current. In this case $S/N({\rm H \beta})/S/N({\rm H \alpha})=2$,
i.e., the H$\beta$ line is potentially easier to observe than the
H$\alpha$ line. Observing the H$\beta$ line should provide a firmer
redshift identification, while the H$\alpha$/H$\beta$ ratio gives
a useful measure of extinction. In figure (\ref{Halpha_fig}), we show
the detection rate of the H$\beta$ line as a function of redshift. 

How can we pick out potential high redshift targets for 
spectroscopy? They may be quickly identified via continuum imaging using the
Lyman-break technique successfully used at low redshifts (Steidel et al 1996). Beyond the redshift of reionization,
an analogous technique using the Gunn-Peterson trough to detect
dropouts can be used. Photometric redshifts may be supplemented by a narrow band
filter search ($R \sim 100-400$) to isolate $H\alpha$ bright galaxies
in a narrow redshift range. The filter width involves some trade-offs:
the filter should be broad enough to cover
a sufficiently large comoving volume, yet narrow enough that the flux density in H$\alpha$ greatly exceeds continuum flux
densities. Note that current plans for NGST allow for multi-object
spectroscopy in the NIR $1-5 \mu$m range, using micro-mirrors but only single
slit spectroscopy in the MIR range, $5-20 \mu$m (Stockman et al
1997). However, even in the NIR a grating spectrograph will be able to
perform spectroscopy for only a few hundred ($\sim$ several percent) of the objects in the
field of view at a given time, thus requiring many pointings to
complete a survey.  Alternatively, a promising proposal is the Infrared Fourier Transform
Spectrograph (IFTS) (Graham et al 1998) which will be able to perform panchromatic
observations in the $1-15 \mu$m range (which is the desired $z<20$
range for H$\alpha$ observations). It will be able to acquire both
broad band imaging data and higher spectral resolution data
(R=1-10000) for all objects in the field of view simultaneously, with
very high (near $100\%$) throughput. Thus, the need for object
preselection is obviated, and such an instrument would be extremely
well suited for our purposes.

\subsection{Comparison with Ly$\alpha$}

While ionizing objects are expected to be more luminous in Ly$\alpha$
than in H$\alpha$, detection in H$\alpha$ has a number of nice
features. Firstly, since H$\alpha$ is emitted in rest-frame optical,
it is less susceptible to attenuation by dust than Ly$\alpha$, or rest-frame UV
continuum. It is thus a significantly
cleaner tracer of the star formation rate. Furthermore, the
resonant scattering of Ly$\alpha$ photons within the HII region and in
surrounding HI regions means that they will diffuse over a large
region, resulting in an unobservably low surface
brightness. The large optical path of Ly$\alpha$ photons
as they random walk out of these regions greatly increases their
probability of absorption by dust (Charlot \& Fall 1993). Finally, the observability of
Ly$\alpha$ is affected by complex velocity flows of the neutral gas,
which may be outflowing from the ionizing source (thus, for instance,
the observation of P-Cygni type profiles in local star-forming
galaxies (see Kunth el al 1998 and references therein)). Note that
50\% of the objects selected by the Lyman-break technique
show no emission in Ly$\alpha$, while the rest have weak rest-frame
equivalent widths $< 20 \, {\rm \AA}$ (Steidel et al 1996). By
comparison, the expected rest-frame equivalent widths of star-forming
galaxies from modelling their spectral energy distribution is ${\rm
W_{\alpha}} \sim 100-200 \, {\rm \AA}$ (Charlot \&
Fall 1993). Also, note that quasars are expected to have equivalent widths
in the quasar rest frame of ${\rm
W_{\alpha}}= 0.68 \frac{c}{\nu_{\alpha}f_{c}} \int^{\infty}_{\nu_{L}}
d\nu f_{\nu}/\nu
\approx 827 \alpha^{-1} (3/4)^{\alpha}  {\rm \AA} = 360
{\rm \AA} (\alpha = 1.5)$ (Charlot \& Fall 1993), where $\nu_{L},
\nu_{\alpha}$ are the Lyman limit and Ly$\alpha$ frequencies, $f_{c}$
is the continuum flux near Ly$\alpha$, and $f_{\nu} \propto \nu^{-\alpha}$
blueward of Ly$\alpha$. In fact, most bright quasars are observed to have
rest-frame equivalent widths of $50
\leq {\rm W_{\alpha}} \le 150 {\rm \AA}$  (Schneider, Schmidt \& Gunn
1991, Baldwin, Wampler \& Gaskell 1989). This could be due to 
covering of the AGN source by HI clouds or dust extinction effects. We tentatively conclude that
even in the post-reionization epoch, a Ly$\alpha$ survey is an
inefficient means to search for ionizing sources. Until recently,
H$\alpha$ surveys of high-z objects have not been carried out due to the difficulties of
infrared spectroscopy from the ground. It is intriguing to note that recent such H$\alpha$ surveys 
(Glazebrook et al 1998) show a star formation rate approximately three
times as high as that inferred from the 2800 ${\rm \AA}$ continuum
luminosity by Madau et al (1996). This reinforces the expectation that
surveys in UV continuum and Ly$\alpha$ tend to underestimate the star
formation rates as they are subject to highly uncertain dust
extinction effects.

Most importantly for our purposes, a H$\alpha$ survey would be able to
detect ionizing sources prior to the epoch when HII regions overlap,
when the universe was optically thick to Ly$\alpha$ photons. In order
for Ly$\alpha$ emission from a source to be unaffected by resonant scattering, the ionizing source must ionize a region large
enough for the Ly$\alpha$ photon to redshift out of the Gunn-Peterson
damping wing. This problem has been considered by Miralda-Escude
(1998). Given that the optical depth of the Gunn-Peterson damping
wing as a function of displacement $\Delta \lambda$ from line center, $\tau(\Delta \lambda)= 1.7 \times 10^{-3}
(1+z/10)^{3/2} (\Delta \lambda / \lambda)^{-1}$, a
region $r \sim 1$ Mpc in proper size must be ionized, independent of redshift, in order to have
$\tau(\Delta \lambda) \sim 0.3$. Note that this
is somewhat greater than the proper radius of influence of a single ionizing source at
z=9, $r_{\rm influence} \sim 0.1 {\rm Mpc} (M/10^{9} M_{\odot})^{1/3}
(1+z/10)^{-2}$ (assuming the source never reaches its equilibrium
Stromgren radius). Nonetheless, the clustering of ionizing sources means
that many should exist in large, overlapping ionized regions. Even if
an ionized region $r \sim 1 \, $Mpc is created, the high recombination rate
due to the higher IGM densities are high redshift results in a
residual neutral fraction which is optically thick to Ly$\alpha$.  
If we take the column density to be $N_{HI} \sim x n_{b} r$, where $x$
is the ionization fraction in ionization equilibrium, then the optical
depth to Ly$\alpha$ scattering in transversing this region is:
\begin{equation}
\tau_{o} \simeq 100 \left( \frac{1+z}{10} \right)^{6} (1+
\delta_{b})^{2} \left( \frac{J_{21}}{0.5} \right) \left
( \frac{T}{10^{4} K} \right)^{-0.7} 
\end{equation}
where $J_{21}$ is the background ionizing intensity in units of
$10^{-21} {\rm erg \, s^{-1} \, cm^{-2} \, Hz^{-1} \, sr^{-1}}$, and
$\delta_{b}=\delta \rho_{b}/\rho_{b}$ is the gas overdensity. It therefore seems dubious that unscattered Ly$\alpha$ can be observed except in
highly underdense regions, or unless ionizing sources are considerably
more luminous than we have considered. 

Loeb \& Rybicki (1999) have considered the observational signature of
Ly$\alpha$ emission of an ionizing halo, given that the Ly$\alpha$
photons are likely to resonantly scatter in the surrounding neutral
IGM, before redshifting out of resonance. They find that the Ly$\alpha$ photons emitted by a typical source
at $z \sim 10$ scatter over a characteristic angular radius of $\sim
15''$ around the source and compose a line which is broadened and
redshifted by $\sim 10^{3} {\rm km \, s^{-1}}$. However, the large angular scale
leads to a very
low surface brightness. Adopting the same estimates as the previous
section for instrumental noise and the IR background, we find that the
signal is likely to be overwhelmed by instrumental noise. Assuming $\dot{N}_{\rm Ly \alpha} \approx
\dot{N}_{ion}$, so that the flux $f_{\rm Ly \alpha} \approx 12 f_{\rm H
\alpha}$ and the flux density $j_{\rm Ly \alpha} \approx 2.2 j_{\rm H
\alpha}$ (Case B), the relative signal to noise ratio at the same spectral resolution
is given by:
\begin{equation}
\frac{\frac{\rm S}{\rm N}({\rm Ly} \alpha)}{\frac{\rm S}{\rm N}({\rm
H \alpha})}(z<6.6,z>6.6)= (8 \times 10^{-3} ,0.1) \left(\frac{\Omega_{\rm
H_{\alpha}}}{0.01 {\rm arcsec^{2}}} \right)^{1/2} \left(\frac{\Omega_{\rm
Ly_{\alpha}}}{700 {\rm arcsec^{2}}} \right)^{-1/2} 
\label{compare_ly_h}
\end{equation}
(in fact, line broadening for Ly$\alpha$ is likely to further lower
this ratio by a factor of a few). Thus, with $ {\rm S/N(Ly \alpha)}
= 6 \times 10^{-3} (\Omega/700 {\rm arcsec^{2}})^{-1/2} \left( \frac{1+z}{10} \right)^{-1} \left( \frac{\rm M}
{10^{9} {\rm M_{\odot}}} \right)
t^{1/2}$, our canonical source $10^{9} {\rm M_{\odot}}$ source at z=9 would require an unacceptably long
integration time of 32 days for a 10$\sigma$ detection. If even trace
amounts of dust are present, the large number of resonant scatterings
means that the Ly$\alpha$ signal is likely to be severely
attenuated. We conclude that this would be a very difficult
observation, although extremely interesting if carried out
successfully. In particular, it would be very interesting to
view the same bright source in Ly$\alpha$ and H$\alpha$. Their
relative intensities would establish the importance of dust extinction, while their relative
spatial extent would constrain the topology of neutral gas surrounding
the source. Note that Loeb \& Rybicki (1999) assume a uniform IGM surrounding
the source. If the surrounding IGM is inhomogeneous, the Ly$\alpha$
photons are likely to diffuse along the underdense directions, leading
to a signal of even lower surface brightness. In addition, Lyman limit
systems will absorb a certain fraction of the Ly$\alpha$ photons
(since photons scattering in these systems are no longer redshifting
out of resonance).

\section{Free-free emission}

Free-free emission from
galaxies forms a fluctuating background in radio frequencies similar
to the EBL sought in the optical (Vogeley 1997) and the CIB sought after in
the infra-red (Kashlinsky et al 1996, Dwek et al 1998). It is has the potential to
yield a great deal of useful information in that it is a direct tracer of star formation and
clumping, it is unaffected by dust, there are no unknown K-corrections
to be applied (due to the flat spectrum nature of free-free emission), and sources at very high redshifts (z $>$ 10) make a
significant contribution and are still potentially observable. We adopt two
approaches in characterizing the Free-Free Background (FFB): detecting
the mean signal (via spectral distortion of the CMB) and detecting the fluctuating
signal (by direct imaging of ionized halos or detecting temperature
fluctuations across the sky).

\subsection{Spectral distortion}

Consider first the mean sky averaged signal $\langle S \rangle$. The
integrated emission from ionized halos creates a spectral distortion
of the CMB which is potentially observable. Note that for a given mean
surface brightness $\langle S \rangle$, the temperature perturbation
rises quadratically at low frequencies, where (in the Rayleigh-Jeans
limit) $\Delta T_{ff}=c^{2}\langle S \rangle /2k\nu^{2}$. This spectral distortion may be characterised by the optical depth to free-free emission (Bartlett \& Stebbins 1991) $Y_{ff}=\frac{\Delta T_{ff}}{T_{\gamma}} x^{2}$
where $T_{\gamma}$ is the undistorted CMB photon temperature, $x
\equiv h\nu/kT_{\gamma}$ and $\Delta T_{ff}=T_{\rm eff}-T_{\gamma}$ is
the mean temperature perturbation created by free-free emission (note
that the $Y_{ff}$ has a different spectral form from Compton y-distortion (parameterized by $y$), arising from Compton scattering of CMB photons
off hot gas). The quadratic nature of the distortion means that a bound on $Y_{ff}$ is 
best obtained by comparing temperature measurements at high
frequencies (when the perturbation due to free-free emission is small) and low
frequencies (when it becomes important). The CMB blackbody
temperature near the peak of emission has been measured by the FIRAS
instrument aboard COBE to be $T_{\gamma}= 2.768 \pm 0.004$K (95\%
CL) (Fixsen et al 1996), while at low frequencies the effective temperature
has been measured as $T_{\rm eff}=2.65^{+0.33}_{-0.30}$K at 1.4
GHz (Staggs et al 1996a) and $T_{\rm eff}=2.730 \pm 0.014$K at 10.7 GHz
(Staggs et al 1996b). Combining the limits on extant data to
compute $\Delta T_{ff}$, one obtains $Y_{ff} < 1.9 \times 10^{-5}$
(95\% CL) (Smoot \& Scott 1996). Significant amounts of free-free emission can still be present without violating this constraint. At low frequencies, it corresponds to a limit on the mean temperature perturbation: 
\begin{equation}
\Delta T_{ff}\le 4.4 \times 10^{-2} \left(\frac{\nu}{2 GHz} \right)^{-2}
\ K
\label{ydistort}
\end{equation}

One of the chief goals of the DIMES satellite (Kogut 1996) is
to measure spectral distortions from free-free emission at low
frequencies. DIMES will have a frequency coverage 2--100 GHz and a
sensitivity $0.1$mK (at 2 GHz). It is hoped that a measurement of $\Delta
T_{ff}$ will indicate the optical depth of the ionized IGM and thus
constrain the redshift of reionization $z_{r}$. However, we have
already seen from quite general arguments in section 2.2 that the mean
emission from ionized halos should dominate that of the IGM. One can
examine this in more detail. The equation of cosmological radiative transfer (Peebles 1993) gives the
observed surface brightness of the ionized IGM as:
\begin{equation}
J(\nu_{o}) = \frac{1}{4 \pi} \int_{0}^{z_{r}} dz
\frac{dl}{dz} \frac{\epsilon(\nu(1+z))}{(1+z)^{3}} 
\label{pathint}
\end{equation}
where $\epsilon$ is given by equation (\ref{free_emissivity}). If
$z_{r}=13$, as in our model, then this translates into $\Delta
T_{ff}=c^{2}J/2k\nu^{2}= 6 \times 10^{-6}$K at 2GHz, which is beyond the 0.1mK sensitivity of
DIMES, which will only be able to detect $z_{r} > 85$ for the $\Lambda$CDM cosmology we have adopted. On the other hand, if we compute $\langle S \rangle$ from
equation (\ref{moments}) (setting $z_{\rm min},S_{c}=0$ since point
source removal is not possible with the $10^{\circ}$ beam of DIMES),
we obtain $\Delta T_{ff}= 3.4 \times 10^{-3}$K, well within the
capability of DIMES, while still satisfying the y-distortion constraint
by an order of magnitude. Note that we have assumed that free-free
emitters are the only radio-bright sources in the sky. The existence
of radio-loud galaxies and AGNs(see section 4.3.3 for discussion) is likely to
increase the spectral distortion due to radio sources. 

Thus, it is likely that any spectral distortion detected by DIMES will
not constrain the reionization of the IGM but rather the integrated free-free
and other radio emission from galaxies, which provides a stronger
signal by at least an order of magnitude. 

\subsection{Direct Detection of Ionized Halos}

A natural way to determine if an observed spectral distortion is due
to free-free emission from the first ionized halos or the IGM is to
attempt to directly detect sources with a high angular
resolution instrument. 

\subsubsection{Detection of point sources}

The free-free flux from an ionized halo may be estimated as:
\begin{equation}
S_{\rm ff}= \frac{L_{\nu}^{\rm ff}}{4 \pi d_{L}^{2}}
(1+z) \approx 2.5 \left( \frac{1+z}{10} \right)^{-1}  \left( \frac{\rm M} {10^{9} {\rm M_{\odot}}}
\right) \left( \frac{\rm T} {10^{4} \, {\rm K}} \right)^{-0.35} {\rm nJy}
\label{J_free_free}
\end{equation}
where we have estimated $L_{\nu}^{\rm ff}$ from equations (\ref{L_free})
and (\ref{Ndot_scaling}). Assuming a point source, and the 0.1''
resolution of the Square Kilometer Array (SKA), this translates into a
brightness temperature of:
\begin{equation}
T_{b}^{ff}= 0.09 {\rm K} \left( \frac{\nu}{2 \, {\rm GHz}} \right)^{-2.1}
\left( \frac{T} {10^{4} {\rm K}} \right)^{-0.35} 
\left( \frac {M}{10^{9} M_{\odot}} \right)
\left( \frac{1+z}{10} \right)^{-1} \left(\frac{\theta}{0.1''} \right)^{-2}
\end{equation}
On the other hand, the detector noise may be estimated as (Rohlfs \&
Wilson 1996): 
\begin{equation}
S_{\rm instrum}= \frac{2 kT_{\rm sys}}{A_{\rm eff} \sqrt{2 t \Delta \nu}}= 10
\left(\frac{\Delta \nu}{1 \, {\rm GHz}} \right)^{-1/2} \left( \frac{t}
{10^{5} \, s} \right)^{-1/2} {\rm nJy}
\label{noise}
\end{equation}
where we have used $A_{\rm eff}/T_{\rm sys}=2 \times 10^{8} {\rm
cm^{2}/K}$ for the SKA (Braun et al 1998), and we assume a bandwidth
$\Delta \nu \approx 0.5 \nu$. Thus, in 10 days, a
5$\sigma$ detection of a 16 nJy source is possible. This is a
tremendous leap in sensitivity; the deepest observations with the VLA at
8.4 GHz (Partridge et al, 1997) with approximately a week's total
integration time were able to identify point sources at
the 7 $\mu$Jy level (the rms sensitivity was $1.5
\mu$Jy), with a resolution of 6''. 

\begin{figure}
\epsscale{1.00}
\plotone{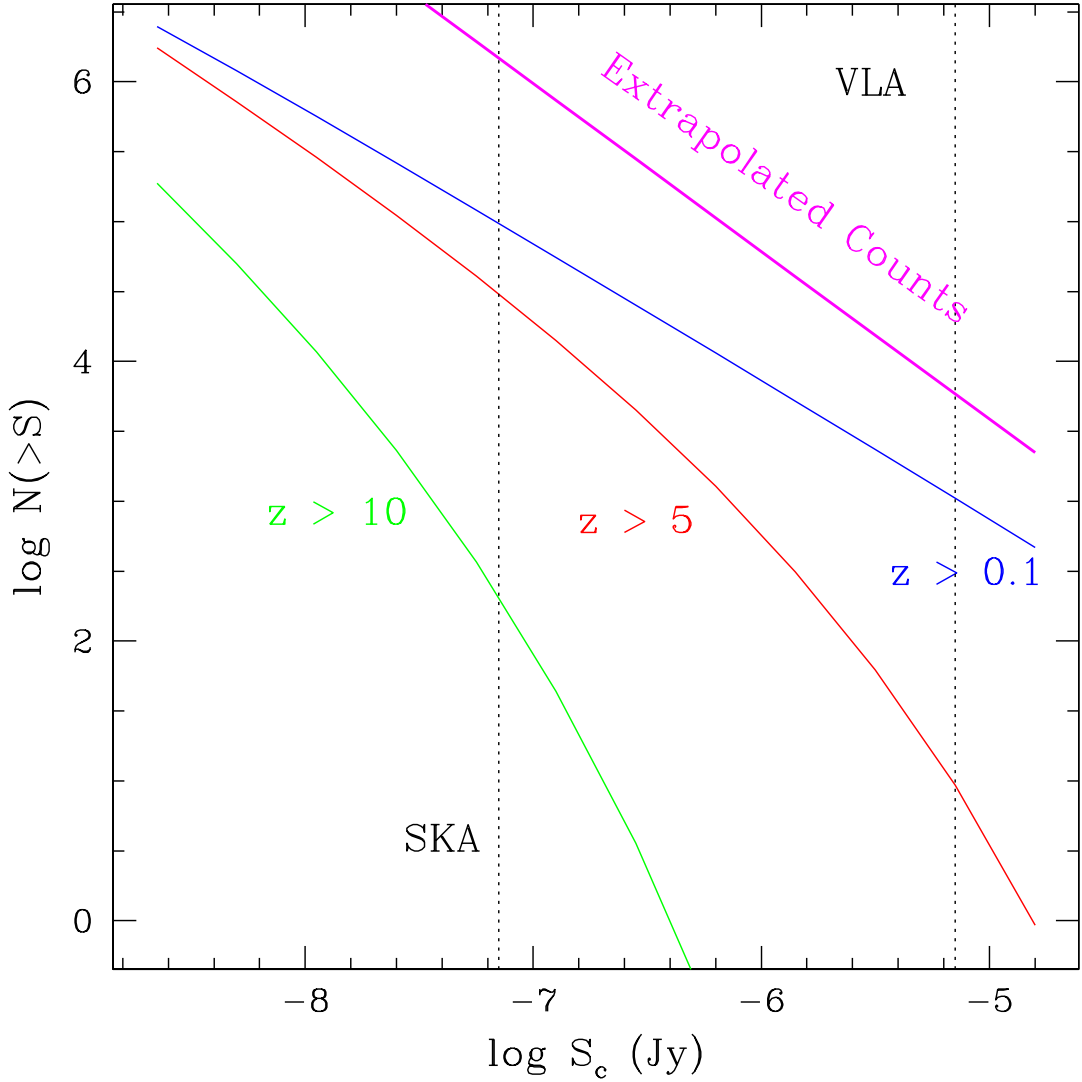}
\caption{Number of sources which may be detected in the $1^{\circ}$
field of view of the Square Kilometer Array (the field of view for the
VLA is similiar, $40' \times 40'$), as a function of the cut-off flux
$S_{c}$. Realistic limiting fluxes for point source detection are shown. The
extrapolated source counts from Partridge et al (1997) are also shown.}
\label{Num_sources_free}
\end{figure}

For a given integration time, the sensitivity for
detection of a given halo in free-free emission with the SKA is
significantly less than that for H$\alpha$ detection (with $R> 1000$) with NGST. The relative signal to noise is:
\begin{equation}
\frac{\frac{\rm S}{\rm N}({\rm free-free)}}{\frac{\rm S}{\rm N}({\rm
H \alpha})}(z<7,z>7) =(1.1 \times 10^{-3},1.3 \times 10^{-2}) \left( \frac{\Delta \nu}{1 \, {\rm GHz}} \right)^{1/2}
\left( \frac{t_{\rm ff}}{t_{\rm H\alpha}} \right)^{1/2} 
\label{compare_free_h}
\end{equation}
where the above ratio holds in the regime where free-free detection is limited by instrumental noise (for a discussion of confusion noise, see section (\ref{foreground_section})). Thus, the SKA will only be able to detect bright sources. However,
note that its field of view is considerably larger than that of
NGST--$1^{\circ}$ rather than 4'. In figure (\ref{Num_sources_free}) we compute
the number of sources detectable above a given flux threshold
$S_{c}$. We find that $\sim 10^{4}$ sources with $z>5$ should be present
in the 1 square degree field of view of the Square Kilometer
Array above a source detection threshold of $70$nJy.

\subsubsection{Power spectrum of unresolved sources}

\begin{figure}
\epsscale{1.00}
\plotone{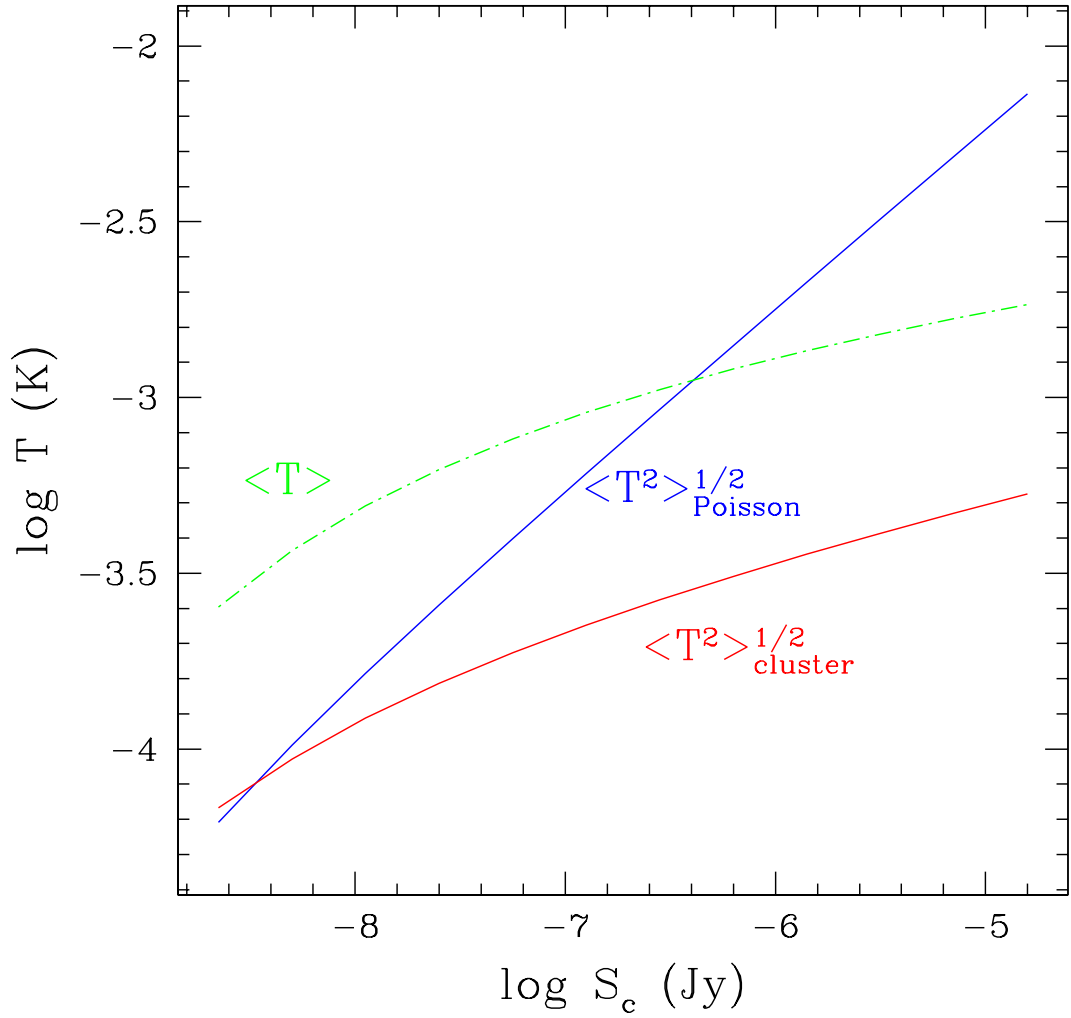}
\caption{Mean and RMS temperature fluctuations for diffuse background
below point source removal threshold $S_{c}$, at 2 GHz and 4''
resolution. Note the steeper dependence of Poisson fluctuations on
the flux cutoff $S_{c}$ compared to the clustering term.} 
\label{background}
\end{figure}

Identification of individual sources may not always be
feasible. Many sources will too faint to be individually detected, or they
may be crowded and blended together in a confused field. Such sources
can still be detected statistically by observing the fluctuations in the FFB
due to sources below the flux limit. The discreteness of sources
create large amounts of small scale power. The power spectrum is
white noise ($C_{l} \sim$ constant) until the angular scale of the beam (or the typical scale
of a halo, if it is larger), beyond
which it damps rapidly. We can compute compute the statistics of the background
below the flux cut $S_{c}$, assuming the sources are unclustered. In particular, we compute the power
spectrum of the fluctuations from equation (\ref{moments}):
\begin{equation} 
C_{l}^{\rm Poisson}=\langle S^{2} \rangle=
\int^{S_{c}}_{0} \frac{\partial N}{\partial S} S^{2} dS
\end{equation}
(independent
of $l$), and thus
compute the rms temperature fluctuations from $ T_{\rm rms}^{2}= \sum_{l} \frac{2l+1}{4\pi} C_{l}^{\rm Poisson}
B_{l}^{2} \approx \frac{C_{l}^{\rm Poisson}}{4 \pi \theta^{2}}$
where we have taken the beam of the telescope to be
Gaussian: $B_{l}^{2}=e^{-l(l+1)\theta^{2}}$. Note that the rms
temperature fluctuations increase with increasing resolution. In
Figure (\ref{background}) we display the computed rms temperature fluctuations
at 2 GHz and with $4''$ resolution. Also
shown is the mean signal $\bar{T}$, computed from the first moment
of equation (\ref{moments}), which is independent of angular
resolution. Our model is consistent with present observations. With $S_{c}=7 \mu$Jy and $\theta=6''$ we obtain $\Delta
T/T= 9.5 \times 10^{-5}$ at 8.4GHz, as compared with the measurement
$\Delta T/T \approx 7 \pm 8 \times 10^{-5}$ of Partridge et al (1997),
and within their 95\% confidence limit $\Delta T/T < 1.3 \times 10^{-4}$.

\begin{figure}
\epsscale{1.00}
\plotone{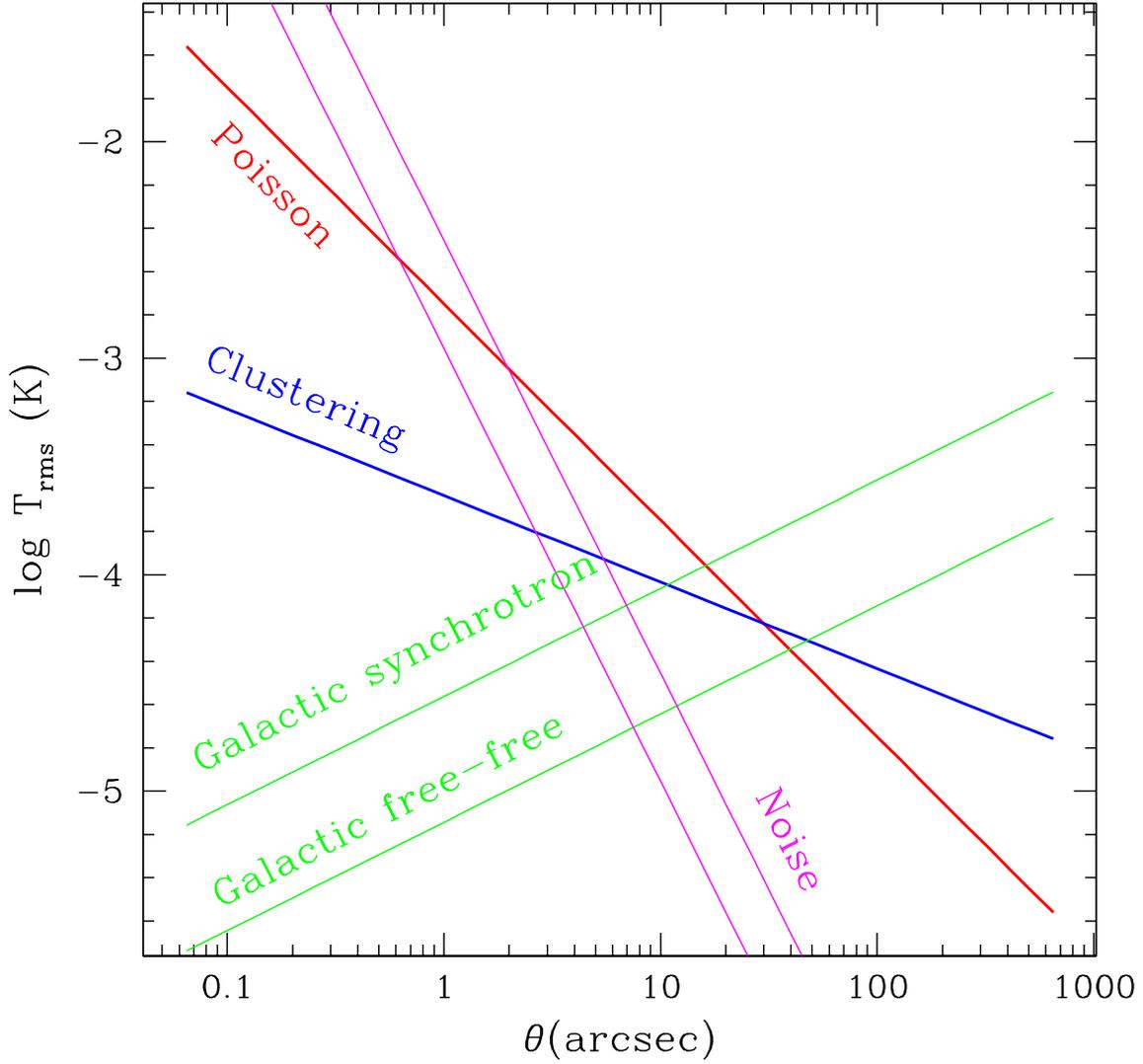}
\caption{Angular dependence of $T_{rms} \equiv [\frac{l(l+1)C_{l}}{4
\pi}]^{1/2}$ at 2 GHz, as a function of the beam FWHM $\theta$. Here the FFB is composed of sources with
$S<S_{c}=70$nJy. Temperature fluctuations of the FFB increase with higher angular
resolution, whereas the foregrounds damp rapidly. Also shown is the
instrumental noise for integration times of 1 and 10 days. There
exists a window in the $2''-20''$ range where the signal is greater
than both instrumental noise and foregrounds. Note that the Poisson
term is greater than the clustering term in this domain.}
\label{angular}
\end{figure}

\begin{figure}
\epsscale{1.00}
\plotone{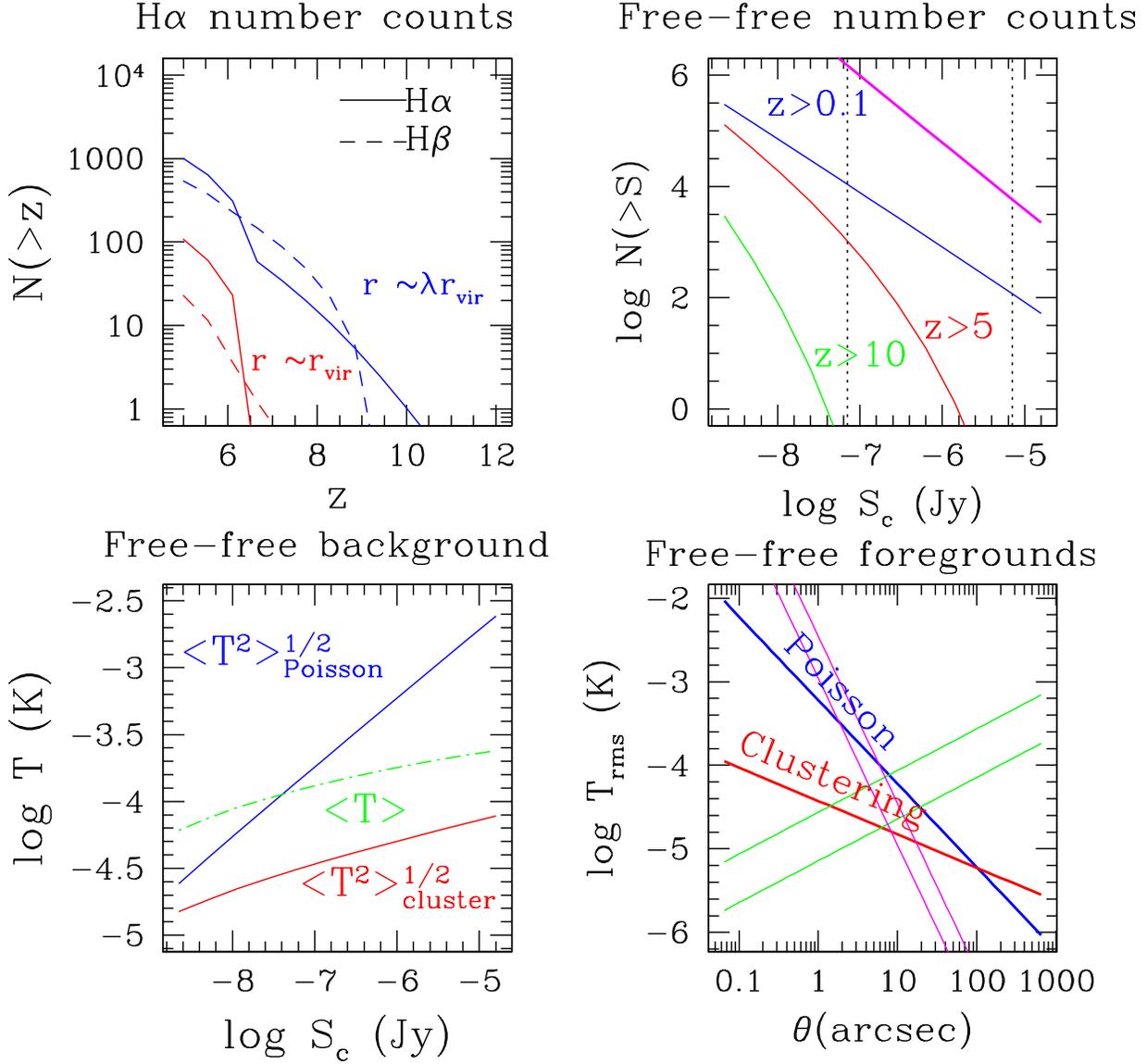}
\caption{Figures (\ref{Halpha_fig}) (H$\alpha$ number
counts),(\ref{Num_sources_free}) (Free-free number counts),
(\ref{background}) (Free-free background),(\ref{angular}) (Free-free
foregrounds), computed for the lower star
formation efficiency case $f_{star}=1.7 \%$. All other assumptions are
unchanged. }
\label{lower_SF_efficiency}
\end{figure}

Asssuming that galactic foregrounds are negligible (see section
(\ref{foreground_section})), the main source of noise is
instrumental. From equation (\ref{noise}), the rms sensitivity to
temperature fluctuations for a beam of angular size $\theta$ is:
\begin{equation}
T_{noise}=0.35 \, {\rm K}\left(\frac{\nu} {2 \,{\rm GHz}} \right)^{-2.1}
\left(\frac{\theta}{0.1''} \right)^{-2} \left(\frac{\Delta \nu}{1 \, {\rm GHz}} \right)^{-1/2} \left( \frac{t}
{10^{5} \, s} \right)^{-1/2}
\end{equation} 
For the same parameters, and assuming a cut-off flux of $S_{c}=70$nJy,
our expected signal is only $0.02$K. However, note that our expected signal $T_{rms}
\propto \theta^{-1}$, whereas the instrumental noise $T_{\rm instrum}
\propto \theta^{-2}$. By varying the size of the interferometer,
we can vary the angular size of the beam, optimising it for this
observation. Referring to figure (\ref{angular}), we see that
on angular scales $\sim 1''-10''$, the temperature fluctuations due to
the free-free background exceeds both the instrumental noise level and
the galactic synchrotron foreground. Therefore, observing a radio
field at coarser resolution increases the signal to noise ratio of
temperature fluctuations sufficiently to enable the fluctuating
free-free background to be detected. Note that if we simply smooth the
map, the $T_{\rm instrum}
\propto \theta^{-2}$ scaling is not strictly applicable. Smoothing the map corresponds to downweighting longer baseline
information, which essentially means that some information is thrown
away. This decreases the effective area and increases the detector
noise, so $T_{\rm instrum}$ declines more gradually with $\theta$
(although still faster than $T_{rms}
\propto \theta^{-1}$). The best solution would be to
observe in a more compact configuration with resolution $\sim
1''-10''$ to begin with, so no data is thrown away. One wants to reach
a compromise between the need for the high angular resolution required to
beat down source confusion in discrete source identification and
removal, and the somewhat lower angular
resolution optimal for analysing the fluctuating residual
signal, or confusion noise itself. A natural solution would be to work
at an angular resolution $\theta_{\rm equal}$
where instrumental and confusion noise are comparable. Working at
higher resolutions than this does not yield significant advantages for
discrete source extraction, since the instrumental noise background in
flux units is independent of angular resolution. By fiat, after
source removal the
instrumental noise and fluctuating signal are now about equal, and
smoothing the fluctuating signal increases the signal to noise ratio
sufficiently for an analysis to be undertaken.  
Note that source removal need not always be in the instrument noise
dominated regime, $S_{c}=n_{\sigma} S_{\rm instrum}$. Source
identification in a crowded field can be difficult, and not all
sources can be removed down to the flux limit. If the angular size
of typical objects is large, $\theta_{\rm typ} > \theta_{\rm equal}$,
then confusion noise exceeds instrumental noise and $S_{c} >
n_{\sigma} S_{\rm instrum}$. In this case, increasing
the angular resolution beyond $\theta_{\rm typ}$
yields no advantage. Both because of the higher cut-off flux $S_{c}$,
as well as the lower angular resolution, detecting the fluctuating signal will not be a
problem. Furthermore, as we discuss in
section (\ref{foreground_section}), our model probably
underestimates the temperature fluctuations which will be observed,
which is likely to have components from other radio souces in the sky
such as low-redshift radio galaxies and radio-loud AGNs. 

The free-free component may be
separated from other contaminants in multi-frequency observations by
its spectral signature (flat spectrum, as opposed (for instance) to the
$B_{\nu} \propto \nu^{-0.9}$ spectrum of synchrotron
radiation). In addition, besides merely computing RMS temperature
fluctuations, it would be desirable to compute the angular power
spectrum of the observed signal and
confirm its white noise character. In performing this analysis one has
to take care to ensure that there is no aliasing of power on scales
larger than the field of view (due to Galactic foregrounds and the
CMB). This can be done by high-pass filtering and edge tapering the
map. Besides the various sources of
noise depicted in figure (\ref{angular}), one also has to contend with
sample variance, given by:
\begin{equation}
(\Delta C_{l})^{2}= \frac{1}{f_{\rm sky}} \frac{2}{2 l+1} C_{l}^{2}
\label{sample_variance}
\end{equation}
where $f_{\rm sky}$ is the fraction of sky covered (this expression
assumes the $C_{l}$'s are Gaussian distributed, as they should be
by the Central Limit Theorem, since a large number of modes contribute). In addition, one probably will bin the
computed $C_{l}$'s in bins of size $\Delta l$ to increase the signal
to noise. Using $l \, \theta \sim \pi$, an angular resolution of $\theta \sim
4''$ corresponds to $l \sim 10^{5}$. The signal to noise for the
sample variance term is $\frac{C_{l}}{\Delta C_{l}}=20 
\left( \frac{l}{10^{5}} \right)^{1/2} \left(\frac{\Omega}{1^{\circ} \,
\Box} \right)^{1/2} \left(\frac{\Delta l}{100} \right)^{1/2}$, which
may be increased by increasing the bin size $\Delta l$. Thus, sample
variance should not pose any problems. 

\subsubsection{Clustering of ionized sources}

Does clustering of the ionized halos make a significant contribution
to fluctuations in the FFB? We find that on arcsecond scales and with a
cutoff flux $S_{c}=70$nJy, Poisson fluctuations are more important
than the contribution due to clustering. However, with a lower cutoff
flux $S_{c}$, or on somewhat larger angular scales, the clustering term
dominates. 

\begin{figure}
\epsscale{1.00}
\plotone{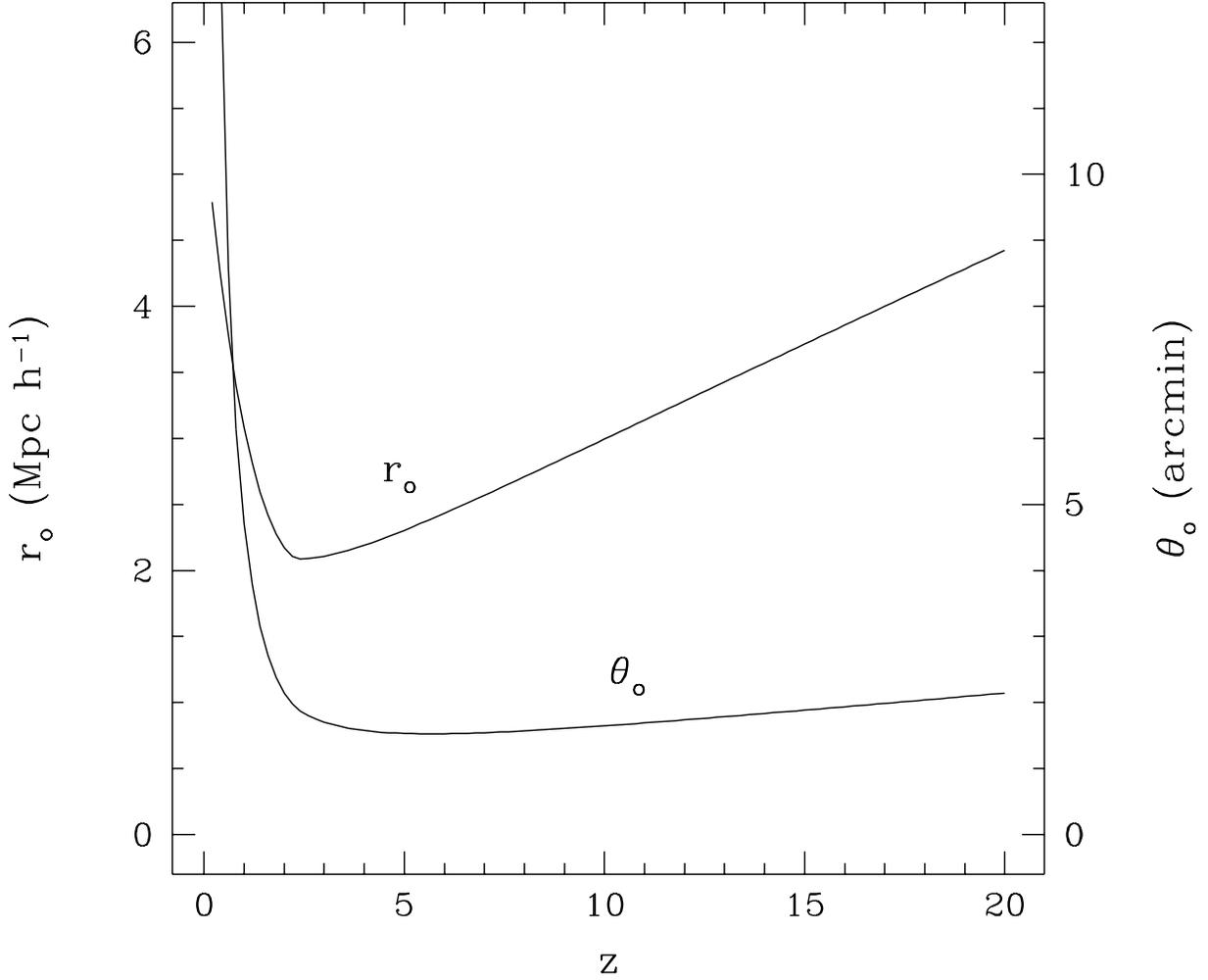}
\caption{Characteristic number weighted correlation length $r_{o}$
(comoving) as
function of redshift, and corresponding angular scale $\theta_{o}
\equiv r_{o}^{\rm proper}/d_A$. The bias of
high redshift halos means that angular clustering does not plummet
with redshift. For a higher cutoff mass $M_{*}$, bias becomes important 
at lower redshift. Beyond the turnoff, the angular scale of clustering
is almost constant at a few arcminutes.}
\label{cluster}
\end{figure}

First, let us consider the evolution of the
halo correlation function with redshift. The matter correlation
function may be computed as:
\begin{equation}
\xi_{mm}(r,z) = \int_{0}^{\infty} \frac{dk}{k} 4 \pi k^{3} P(k,z) \frac {sin \
kr} {kr} 
\label{correl_func}
\end{equation}
However, halos which collapse early in the history of the universe
consitute rare density peaks which are highly biased. In a linear bias
model, the correlation function of halos is given by:
\begin{equation}
\xi_{hh}(M_{1},M_{2},r,z)=b(M_{1},z)b(M_{2},z) \xi_{mm}(r,z)
\end{equation}
where the bias factor is given by Mo \& White (1996):
\begin{equation}
b(M,z)=1+ \frac{\nu^{2}-1}{\delta_{c}} = 1+ \frac
{\delta_{c}}{\sigma^{2}(M,z)} - \frac{1}{\delta_{c}}
\label{bias}
\end{equation}
where $\delta_{c}=1.68$ and $\nu=\delta_{c}/\sigma(M,z)$. For
simplicity we assume the correlation function of halos has the same shape at
different redshifts, and differs only in amplitude. This assumption is
shown to hold in numerical simulations (Kravtsov \& Klypin 1998, Ma 1999), due
to a cancellation of non-linear effects in the evolution of the power
spectrum and bias factor. We assume $\xi_{mm}(r)$ is a power law on
scales below the non-linear lengthscale (since this is shown to hold
in numerical simulations), and is given by equation
(\ref{correl_func}) on scales above the non-linear lengthscale. In Figure (\ref{cluster}), we compute the
halo correlation
length $r_{o}^{halo}$ (defined as $\xi_{hh}(r_{o}^{halo},z)=1$) at each epoch $z$, assuming a number weighted bias factor
\begin{equation}
\tilde{b}(z)= \frac {\int_{M_{*}}^{\infty} b(M,z) n(M,z) dM} {\int
_{M_{*}}^{\infty} n(M,z) dM}
\end{equation}
The cutoff mass $M_{*}=10^{8} (1+z/10)^{-3/2} M_{\odot}$ introduces strong bias at high redshift, with
the net result that the correlation length of halos decreases very
slowly with redshift, despite the fall in the matter correlation
length. For a higher cutoff mass, bias becomes important at lower
redshift, and the initial dip seen in Figure (\ref{cluster}) moves to
lower redshift and is less pronounced. Thus, the observed
clustering in the Lyman-break galaxy sample of
Giavalisco et al (1998), with a comoving correlation length at $z \sim
3$ comparable
to that in the local universe, is consistent with a detection of the most
luminous and massive ($M \sim 8 \times 10^{11} h^{-1} M_{\odot}$)
galaxies at that redshift. The clustering of ionising sources at high redshift is therefore
non-negligible, and a potentially important effect in contributing to
free-free background fluctuations.  

The angular correlation function of the free-free background below a
flux cut-off $S_{c}$ is:
\begin{eqnarray}
C(\theta,S_{c})&=& \left( \frac{1}{4 \pi} \right)^{2} \int_{0}^{\infty} dz
\frac{dl}{dz} \frac{j_{eff}^{2}(z,S_{c})}{(1+z)^{6}} \int_{-\infty}^{\infty}
d\Delta \xi_{hh}(r,z) \\ \nonumber
&=& \left( \frac{1}{4 \pi} \right)^{2} \int_{0}^{\infty} dz
\frac{dl}{dz} \left[ \int_{S_{min}(z)}^{S_{c}}
dS \frac{dN}{dVdS} \left( \frac{4
\pi d_{L}^{2} S}{(1+z)} \right) b(S,z) \right]^{2} \
\int_{-\infty}^{\infty}
d\Delta \xi_{mm}(r,z)
\end{eqnarray}
where $j_{eff}(S_{c},z)$ is the proper free-free emissivity of halos at
redshift $z$ below the flux limit $S_{c}$, and
$b(S,z)=b(M(S,z),z)$. We use the small angle approximation $r^{2}=\Delta^{2}
+d_{A}^{2}\theta^{2}$, an excellent approximation since the
clustering length covers a small part of the sky, $r_{o}/d_{A} \ll 1$.
The contribution to the angular power spectrum due to clustering
$C_{l}^{\rm cluster}$ is then straightforwardly computed from the
Legendre transform of the correlation function:
\begin{equation}
C_{l}^{\rm cluster}= 2 \pi \int d(cos \theta) P_{l}(cos \theta)
C(\theta) \approx 2 \pi \int_{0}^{\infty} \theta d\theta J_{o}(l\theta)
C(\theta) 
\end{equation}
where the last step is valid for $l \gg 1$. 

In Figure (\ref{background}), we show the dependence of the clustering
term on the flux cutoff at fixed angular scale $\theta \sim 1''$. As
we lower the flux cutoff $S_{c}$, the Poisson term declines more
sharply than the clustering term, until the clustering term
dominates. We can understand this intuitively: the Poisson
term is dominated by rare bright objects just below the detection
threshold. As we lower the detection threshold, such objects are
excluded from the residual FFB, leaving a more uniform distribution of
faint objects. By contrast, the clustering term, like the mean
brightness term, is dominated by faint,
low mass objects. In particular, the amplitude of the clustering term is
determined by the bias of objects at the mass cut-off, $M_{c}=10^{8}
(1+z/10)^{-3/2} \, M_{\odot}$, rather than the bias of the rarer (though more
highly biased) high mass objects. Thus, subtraction of bright sources
decreases the Poisson term much more effectively than the clustering
term. In Figure (\ref{angular}), we compute the amplitude of the clustering
contribution to temperature fluctuations as a function of angular
scale at fixed flux cutoff $S_{c}=70$nJy. We find that on the arcsecond scales on which the surface brightness
of free-free fluctations exceeds instrumental noise and Galactic
foregrounds (see below), the Poisson term is more important than the
clustering term. However, at a lower flux cutoff the clustering term
would dominate on these scales. Note the different
angular dependence of the two terms: $C_{l}^{Poisson} \sim $const,
whereas $C_{l}^{clustering} \propto l^{\gamma-3}$, where $\xi_{hh}
\propto r^{-\gamma}$. On large scales, the matter correlation function
falls below a power law, and
eventually $C_{l}^{cluster} \propto l^{n}$ (where the linear power
spectrum $P(k) \propto k^{n}$). This downturn is unimportant on the
scales plotted, though it becomes appreciable on larger scales. At such low fluctuation amplitudes, both galactic
foregrounds and intrinsic CMB anistropies dominate over the free-free
signal.

\subsubsection{Skewness}

Another signature of residual discrete sources is the skewness they
induce in maps (Refregier, Spergel \& Herbig 1998). We compute this as
the third moment of equation (\ref{moments}), $I_{3}= \Omega_{pix}
\int_{0}^{S_{c}} dS \frac{dn}{dS} S^{3} \, {\rm Jy^{3}}$. The instrumental
noise is Gaussian distributed, and thus has vanishing skewness. The
1$\sigma$ measurement uncertainty on this vanishing skewness is
$\sigma[\eta_{instrum}]=\sqrt{\frac{6}{N_{pix}}}
\sigma_{instrum}^{3}$, where $\sigma_{instrum}$ is the rms noise per
pixel. Thus, the signal to noise of a skewness measurement is
$\frac{S}{N}= \left(\frac{N_{pix}}{6} \right)^{1/2} \left(\frac{I_{3}} {\sigma_{instrum}^{3}} \right)$. We assume $\sigma_{instrum} \sim 1 \mu$Jy for the VLA, and $\sigma_{instrum} \sim 10 $nJy for the SKA, and also that point sources may be detected
and removed at $S_{c}=7 \mu$Jy for the VLA and $S_{c}=70$nJy for
the SKA. The number of pixels is given by $N_{pix}=(FOV/\theta_{pix})^{2}$,
where FOV=40' for the VLA and FOV=$1^{\circ}$ for the SKA. We
assume pixel sizes of order $\sim 10''$ for the VLA and $\sim 1''$ for the SKA. From equation (\ref{moments}) we obtain:$I_{3}=(1 \mu{\rm
Jy})^{3}$ for the
VLA and $I_{3}= (12 {\rm nJy})^{3}$ per pixel for the SKA. This
translates into $S/N\approx 5^{3}(\theta/10'') (\sigma_{\rm instrum}/1
\mu Jy)^{3}(S_{c}/7 \, {\rm \mu
Jy})^{(4-\beta)}$ for the VLA and  $S/N\approx 13^{3}(\theta/1'')
(\sigma_{\rm instrum}/10 \,
{\rm nJy})^{3}
(S_{c}/70 {\rm nJy})^{(4-\beta)}$ for the SKA, where $\frac{dn}{dS} \propto S^{-\beta}$ in the neighbourhood of $S_{c}$. The skewness is thus a very
efficient way of detecting the residual field, potentially much more
powerful than attempting to use the second moment. It uses the fact
that free-free sources can only contribute positive temperature
fluctuations, whereas fluctuations due to instrumental noise are
symmetric about zero. It is thus a fairly robust statistic. Depending
on how accurately Gaussian the instrumental noise is, the same
exercise can be carried out with kurtosis. Note that we have assumed that Galactic foregrounds
have no skewness. Given that their second moment damps rapidly at
small scales (see below), both the amplitude and variance of the third
moment is likely to be small. It would be
interesting to look at existing VLA maps for residual skewness.

\subsubsection{Foreground contamination}
\label{foreground_section}

A possible concern is foreground source contamination. Note that
due to the finite thickness of the surface of last scattering, CMB
fluctuations damp rapidly beyond $l \sim 10^{3} (\theta \sim
0.1^{\circ})$. Thus, intrinsic CMB fluctations from the surface of
last scattering contribute negligibly on the scales probed here,
$C_{l}^{CMB} \approx 0$. One can filter out the large scale power with
a filter function $F_{l}=1-e^{l(l+1)\sigma_{filter}^{2}}$, where
$\sigma_{filter} \sim 0.1^{\circ}$. At small scales, the Galactic
foregrounds $C_{l} \propto B(\nu)l^{-3}$ should damp rapidly(Tegmark \&
Efstathiou 1996, Bersanelli et al 1996), although note that the
empirical $l^{-3}$ power law fit has only been observed for $l<300$ (Wright
1998); we assume this behaviour may be extrapolated to smaller scales. Below, we briefly summarise the estimated contributions from
various foregrounds. 

{\bf Galactic Free-free emission} This component is dangerous as its spectral signature is identical to
our signal. However, the large scale component of the Galactic
free-free emission can be removed by cross-correlation with H$\alpha$
maps. For instance, the WhaM survey (Reynolds et al 1995; see http://www.astro.wisc.edu/wham) will map the northern sky at $1^{\circ}$
resolution down to 0.01 Rayleighs (1 Rayleigh=$2.41 \times 10^{-7} {\rm erg \,
s^{-1} cm^{-2} sr^{-1}}$), or $T_{ff}= 17 \mu{\rm K} T_{4}^{-0.35}
\left(\frac{\nu}{2 {\rm GHz}} \right)^{-2.15}$. We assume $B(\nu)
\propto \nu^{-0.15}$ and normalise to the DIRBE analysis of Kogut et
al (1996), which obtained rms fluctuations of 7.1 $\mu K$ at 53
GHz (COBE's finite beam of course causes an underestimate of the rms
fluctuations, but because $C_{l} \propto l^{-3}$ this is a negligible
effect).  This yields: 
\begin{equation}
C_{l}^{free-free}=4.2 \times 10^{-4}
\left( \frac{\nu}{2 {\rm GHz}} \right)^{-4.3} l^{-3} \ K^{2} \ sr
\end{equation}
which translates into $ \delta T^{free-free}= 2.3 \times 10^{-6} \left( \frac{\theta}{0.1''}
\right)^{0.5} \left( \frac{\nu}{2 {\rm GHz}}
\right)^{-2.15}$ K.

{\bf Galactic Synchrotron} We assume $B(\nu) \propto \nu^{-0.9}$ and
normalise to an rms fluctuation of 11 $\mu K$ at 31 GHz, the upper
limit of Kogut et al (1996), as indicated by a cross-correlation
between 408 MHz and 19 GHz emission (de Oliveira-Costa et al
1998). This yields:
\begin{equation}
C_{l}^{syn}=6.1 \times 10^{-3} \left( \frac{\nu}{2 {\rm GHz}}
\right)^{-5.8} l^{-3} \ K^{2} \ sr
\end{equation}
which yields $\delta T^{syn}= 8.6 \times 10^{-6} \left( \frac{\theta}{0.1''}
\right)^{0.5} \left( \frac{\nu}{2 {\rm GHz}} \right)^{-2.9}$ K. The
synchrotron emission may be identified and eliminated by its distinct spectral signature.

{\bf Extragalactic Point Sources} Extragalactic point sources are
essentially the signal we are looking for. In this paper, we predict
the number of sources in a field of view above a certain flux $S_{c}$
which may be directly detected, as well as attempt statistical
characterizations of the fluctuating background signal due to faint
undetected sources. Here we pause to note two facts. Firstly, there
are many other sources in the radio sky (radio galaxies, AGNs, etc) than the free-free
emitters we have attempted to characterize. Secondly, when detecting
point sources, the fluctuating signal due to undetected sources is
also a source of confusion noise. 

How do the figures in our model compare with existing observations?
The deep radio survey by Partridge et al (1997) obtain a maximum
likelihood estimate of the integral source count as:
\begin{equation}
N(\ge S) = (17 \pm 2)\left(\frac{S}{1 \mu{\rm Jy}} \right)^{-1.2 \pm
0.2} {\rm arcmin^{-2}}
\label{source_counts}
\end{equation}  
Assuming that most sources have flat spectra $I_{\nu} \propto
\nu^{\alpha}$, where $\alpha \sim 0$ in this frequency range, we see
from figure (\ref{Num_sources_free}) that our model yields fewer
sources than the
observed number counts at 7$\mu$Jy by a factor of a few. Our admittedly crude model is not meant to be
accurate at low redshift and ignores many contributions to the radio
sky. Our low source counts, together with the fact that we satisfy
present y-distortion bounds by an order of magnitude, indicate that the
model employed is very conservative and probably underestimates the observable
signal. On the other hand, extrapolating the observed source counts to
fainter flux levels clearly overestimates the signal, as the mean flux
$\langle S \rangle$ diverges and violates the y-distortion
constraint. Evidently the source counts must turn over at low flux
levels (in our model, this occurs because of the low mass cutoff
$M_{*}$). To be conservative, we underestimate the signal to noise
ratio of projected observations. Thus, we employ our semi-analytic
model to estimate the fluctuating background signal, whereas when
estimating confusion noise in point source detectation, we use the
extrapolated source counts. 

Note that we are not completely powerless to distinguish against
contaminants, such as low redshift AGNs.  If star formation occurs in
an extended fashion across galactic halos, we should be able to
spatially resolve our target sources, whereas AGNs would only show up
as point sources. In addition, by observing at several frequencies, the
observed spectral index $\alpha$ (where $S_{\nu} \propto
\nu^{-\alpha}$) can be used to diagnose the origin of radio
emission. In particular, it can be used to distinguish between thermal and
non-thermal emission. Finally, as a zeroth order cut, our target high redshift galaxies should be
amongst the faintest objects in a sample, both because of increased
cosmological dimming, and the fact that collapsed objects at high
redshift are intrinsically less massive and luminous. 

We now estimate the confusion noise. Using equation (\ref{source_counts}), we can place an upper bound on the noise from residual undetected
sources from $\langle S^{2} \rangle=
\int^{S_{c}}_{0} \frac{\partial N}{\partial S} S^{2} dS=25.5 \, S^{0.8} \mu{\rm Jy^{2}} \, {\rm arcmin^{-2}}$, which
yields the white noise power spectrum:
\begin{equation}
C_{l}^{PS}=2.4 \times 10^{-15} \left( \frac{S_{c}}{70 {\rm nJy}}
\right)^{0.8} \left( \frac{\nu}{2 \ {\rm GHz}} \right)^{-4 +2\alpha} \
{\rm K^{2} \, sr}
\end{equation}
This gives $\delta T^{PS}= 0.09 {\rm K} \left
( \frac{\theta}{0.1''} \right)^{-1} \left( \frac{S_{c}}{70 {\rm nJy}}
\right)^{0.4} \left( \frac{\nu}{2 {\rm GHz}}
\right)^{-2+\alpha}$, by far the leading contribution to foreground
contamination at these
frequencies and angular scales.  We can
compare the relative contribution of instrumental and confusion noise
by demanding that the source flux cutoff be $n_{\sigma} \sim 7$ times the rms noise. In this
case, using equation (\ref{noise}), we obtain $\frac{S_{confusion}}{S_{instrum}}=1. \left( \frac{\theta}{0.4''}
\right) \left( \frac{n_{\sigma}}{7} \right)^{0.8} \left( \frac{\Delta
\nu}{1 \, {\rm Ghz}} \right)^{0.5} \left( \frac{t}{10^{5} s}
\right)^{0.5} \left( \frac{\nu}{2 \, {\rm GHz}}
\right)^{\alpha}$. In the confusion noise dominated regime, the
appropriate cut-off flux may be determined by directly comparing the flux received from a source at the cut-off flux $S_{c}$ and the background flux below this cut-off:
\begin{equation}
\frac{S}{N} \approx  \frac{S_{c}}{(\langle S^{2} \rangle|_{S_{c}} \theta^{2})^{1/2}} = 6 \left(\frac{S_{c}}{70 \, {\rm nJy}} \right)^{0.6} \left( \frac{\theta}{0.4''} \right)^{-1}  
\end{equation}
Confusion noise thus becomes
the leading source of noise in discrete source identification and
extraction at low angular resolution or if sources are extended rather
than point-like, i.e.,
$\theta ={\rm max(\theta_{beam},\theta_{object})} > 0.4''$. 

Will source crowding be a problem? If we extrapolate the Partridge et
al (1997) counts down to $70$ nJy, then there will be $\sim 400$ sources per square arcmin, or $\sim 9
{\rm arcsec^{2}}$ per source. With a maximum resolution of 0.1'', source
crowding of bright point sources such as disks or mini-quasars should not be a problem. Even if the sources
are extended, the angular extent of typical high redshift $M_{*}$
objects is small, $\theta_{vir} \sim r_{vir}/d_{A} < 1''$.   However, it is important to note that nothing is known about
the appearance of the radio sky below $1 \mu$Jy. It is possible that
a host of extended low redshift, low surface brightness objects may
suddenly surface at these flux levels.

\subsubsection{Lower Star Formation Efficiency Case}
\label{lower_SF_section}

Let us consider the lower star formation efficiency case where only
$1.7\%$, rather than $17\%$, of the gas mass in halos fragments to
form stars. This corresponds to an IGM metallicity at z=3 of $Z \sim
10^{-3} \, Z_{\odot}$, rather than $Z \sim
10^{-2} \, Z_{\odot}$. The net effect of the lower value is that while the number of sources remain the
same, each source is an order of magnitude fainter. Thus, to obtain the same
number counts as previous plots, an integration time 100 times longer
is required. Also, the brightness of the fluctuating background is
reduced, and is equivalent to removing discrete sources an order
of magnitude fainter. The net result is that point source
detection is likely to be instrument noise limited rather than
confusion noise limited. Thus, after discrete source removal, the
fluctuating residual background will be significantly harder to detect. In figure (\ref{lower_SF_efficiency}),
we show results for the lower star formation efficiency, assuming the
same integration time and source removal threshold as before. Note
from the plot of foregrounds that the fluctuating background signal is
only marginally detectable. Besides the RMS signal, the skewness is
also reduced, to $I_{3}= (0.5 \mu {\rm Jy})^{3}$ for the VLA (down
from $I_{3}=(1 \mu{\rm Jy})^{3}$) and $I_{3}= (5.5  {\rm nJy})^{3}$
for the SKA (down from $I_{3}= (12 {\rm nJy})^{3}$). Skewness at this
level is still detectable, with S/N=$2.5^{3}(\theta/10'') (\sigma_{\rm instrum}/1
\mu Jy)^{3}(S_{c}/7 \, {\rm \mu
Jy})^{(4-\beta)}$ (VLA) and S/N=$5^{3}(\theta/1'')
(\sigma_{\rm instrum}/10 \,
{\rm nJy})^{3}
(S_{c}/70 {\rm nJy})^{(4-\beta)} $ (SKA). 

The situation in reality is likely lie between the two
scenarios we have considered, and thus the actual number counts is 
probably bracketed by our calculations. The temperature
fluctuations in the residual field are detectable if discrete source
removal is confusion limited; otherwise, it seems likely that only the
non-gaussian signature of the residual field will yield useful
information.

\section{Kinetic Sunyaev-Zeldovich effect}

Sunyaev-Zeldovich effects arise from the scattering of CMB photons off
moving electrons, which result in brightness fluctuations of the
CMB. They are a very attractive means of detecting
high redshift objects, in part because of the well-known fact that
the surface brightness of SZ effects is independent of redshift. This
is because the increased energy density $(1+z)^{4}$ of the CMB at high
redshift keeps pace with surface brightness dimming effects. Thus,
since the angular size of an object of fixed physical size increases with
redshift beyond $z \sim 2$, the SZ flux from a given object {\it
increases} with redshift. The kinetic effect is
given by  $\frac{\Delta T}{T_{CMB}}=\frac{v}{c} \tau$ while the
thermal effect is given by $(kT/m_{e} c^{2})\tau$, where the optical
depth $\tau= \int n_{e} \sigma_{T}\, dr$. For the photoionized halos we are
considering, typical temperatures are $T \sim 10^{4}$K. Their peculiar velocities are $v_{rms} \approx v_{o} D(z)^{1/2} \, {\rm km \,
s^{-1}}$, where $D(z)$ is the growth factor (normalised to 1 today)
and the characteristic one-dimensional velocity dispersion
of galaxies in the local universe is $v_{o}\sim 600 \, {\rm km
\,s^{-1}}$ (Strauss \& Willick 1995). Thus,
$(v/c)/(kT/m_{e} c^{2}) \sim 400 (1+z/10)^{-1/2}$ and the kinetic effect dominates
over the thermal effect. For an isothermal halo, we obtain
$T^{SZ}_{b} \approx 3 \times 10^{-7} \left( \frac{1+z}{10}
\right)^{2} \left( \frac{M} {10^{9} M_{\odot}} \right)^{1/3} $ K,
similar to the temperature fluctuation produced by a cluster in
kinetic SZ at low redshift (the high redshift halo is $\sim 1000$ times
smaller, but $\sim 1000$ times denser). 

Unfortunately, this is likely to be difficult to detect. The flux from
a halo is related to the temperature shift by:
\begin{equation}
S_{\nu}=\frac{2k^{3}T_{\gamma}^{2}}{h^{2}c^{2}}q(x) \int d\Omega
|\Delta T_{\nu} (\theta)| =
\frac{2k^{3}T_{\gamma}^{3}}{h^{2}c^{2}}q(x) \frac{v}{c} \sigma_{T}
\frac{N_{e}}{d_{A}^{2}}
\label{SZ_flux}
\end{equation}
where $q(x)= x^{4}/4{\rm sinh}^{2}(x/2)$ is the spectral function,
,$x \equiv h\nu/kT_{\gamma}$, $T_{\gamma}=2.7 {\rm K}$ is the
CMB temperature, and $N_{e}=\int n_{e} d^{3}r \approx
M/m_{p}(\Omega_{b}/\Omega_{m})$ is the total number of free electrons
in the system. From the form of the spectral function, the
kinetic SZ effect peaks at $\nu=217$ GHz. This is well beyond the
frequency coverage of the SKA, which has an upper limit of 20 GHz. At
this frequency, and taking $v \sim v_{rms}$, we obtain a flux density $S_{\nu}(20 GHz)=0.24
(\frac{M}{10^{9} M_{\odot}}) (1+z/10)^{1.5}$nJy. Comparing with
equation (\ref{J_free_free}), the SZ flux density at 20 GHz is $\sim 10$ times
less than the free-free flux density. We can maximise the SZ
flux by going to higher frequencies. At the peak frequency,
$S_{\nu}(217 GHz)= 9.6 (\frac{M}{10^{9} M_{\odot}}) (1+z/10)^{1.5}$nJy,
and the kinetic SZ signal dominates over free-free emission. At these
frequencies, the MMA (see http://www.mma.nrao.edu) would be the
premier instrument, with a resolution of 0.1'' at 230 GHz. However,
its rms sensitivity at this frequency is 0.21 mJky/${\rm min}^{0.5}$,
which means that even in a week's integration time it will only be
able to go down to $\mu$Jy levels. We conclude that SZ detection of
high-redshift ionized halos is not possible in the near future.

However, it is worth noting that the SKA and MMA {\it will} be able to
detect the kinetic SZ effect from the HII regions in the IGM blown by
the first luminous sources. If the escape fraction is $f_{esc} \sim
20\%$ for our $10^{9} M_{\odot}$ source, it will be able to create an ionized
region with $N_{e} \sim \dot{N}_{ion} t_{o} \sim 2 \times 10^{67}$ electrons
(using equation (\ref{Ndot_scaling}) and assuming the source lasts for
$t_{o} \sim 10^{7}$yrs). This corresponds to a region $r_{comoving}=(3
N_{e}/4 \pi n_{comoving})^{1/3} \sim 1.2$ Mpc in size (note that
$\dot{N}_{ion}/V_{\rm proper} \alpha_{B} n_{\rm proper}^{2} \sim 10^{-7} \ll
1$, so the effect of recombinations is negligible). Since this is
much smaller than the coherence length of the velocity field, we
assume all parts of the bubble are moving with a uniform peculiar
velocity. From (\ref{SZ_flux}), we find that the flux from such an ionized bubble is
$S_{\nu}(20 {\rm GHz})= 35$nJy (detectable by the SKA, which, using
equation (\ref{noise}) and $A_{eff}/T_{sys}=1 \times 10^{8} {\rm
cm^{2}/K}$ at 20 GHz (Braun et al 1998) has a $S_{instrum}=6$nJy rms
sensitivity) and
$S_{\nu}(217 GHz)=1.4 \, \mu$Jy (detectable by the MMA). However,
along any given line of sight one intersects not one but a multitude
of ionized bubbles. Thus, the signal has to be detected
statistically. Calculating the CMB fluctuations due to inhomogeneous
reionization has been the subject of much recent work (Agahanim et al
1996, Grusinov \& Hu 1998, Knox et al 1998; see also Haiman \& Knox
1999, Natarajan \& Sigurdsson 1999), and is beyond the scope of this paper. However, we make two
observations. Firstly, predictions have focussed on the possibility of
detection by Planck (detection by MAP seems unlikely). The temperature
fluctuation power spectrum is generally white noise with a peak on the
scale of a typical bubble; the induced $\Delta T/T \sim
10^{-6}-10^{-7}$.  For the anistropies to
be detectable by Planck, the bubbles need to have an
unrealistically large size $r_{\rm comoving} \sim 10$Mpc. For bubbles
with $r_{\rm comoving} \sim 1$Mpc, the anisotropies peak on
sub-arcminute scales, beyond the angular resolution of
Planck ($\theta_{\rm FWHM} \sim 5.5'$ at 217 GHz; see
http://astro.estec.esa.nl/Planck/). High resolution ground based instruments such as the SKA or
MMA, which can easily resolve the bubbles, might be better suited to the task. Furthermore, the significantly
larger collecting area of these instruments means that they will be
more sensitive than Planck by about 3 orders of magnitude: with a year
of observing time, the rms sensitivity of Planck at 217 GHz is at best
$\sim 1$mJy. If we require that the sample variance be sufficiently low that $\delta
\equiv C_{l}/ \Delta C_{l} =10$, then using
equation (\ref{sample_variance}), we find that a map of size $\Omega=
2 (\delta/10)^{2} (l/10^{4})^{-1} (\Delta l/10^{2})^{-1}$ square
degrees is sufficient. This is easily achievable with the SKA. Secondly, the signal due to
inhomogeneous reionization (IR) is quite possibly smaller than that due to
the Ostriker-Vishniac effect, which arises from the coupling between
density and velocity fluctuations in a fully reionized medium (Haiman
\& Knox 1999, Jaffe \& Kamionkowski 1998). To
isolate the IR signal, it
would be interesting to cross-correlate observed CMB fluctuations with
the observed distribution of ionizing halos. They could either correlate
positively (if halos blow ionized bubbles around themselves) or negatively
(if underdense voids are reionized first).

\section{Conclusions}

In this paper, we have used a simple model based on Press-Schechter
theory to make observational predictions for the detection of
high redshift galaxies in free-free and H$\alpha$ emission. These
signals are valuable as they directly trace the distribution of
dense ionised gas at high redshift. Furthermore, they are relatively
unaffected by dust or resonant scattering, and their intensities are
directly related to one another. They can thus serve as a very
powerful probe of the dark ages, and constitute promising applications
of the new generation of instruments designed to
probe this epoch, in particular the Next Generation Space Telescope (NGST)
and the Square Kilometer Array (SKA).

A fairly robust prediction of our model is that the integrated
free-free and other radio emission from ionizing sources will swamp the
free-free emission from the ionized IGM by an least an order of
magnitude. Thus, any spectral distortion of the CMB observed
by DIMES at low frequencies will constrain the former quantity. This claim depends only upon the
small escape fraction of ionizing photons from sources, as is
observed in galaxies in the local universe. To evade it,
the ionizing sources must themselves contain very
little ionized gas, either because gas has been driven out by
supernovae, or ionizing photons are able to escape through a hole in
the ISM, leaving most of the gas in the host galaxy neutral. 

The escape fraction of ionizing photons $f_{esc}$ from high redshift halos may
be constrained by comparing the rest frame UV
luminosities longward of the Lyman limit of the sources as observed by
NGST, with the expected luminosities in H$\alpha$ and free-free
emission. Given a model for the relative spectral energy distribution,
the relative luminosities then directly constrain the escape fraction in halos at high redshift, an important quantity in theories
on reionization. If this fraction is low, as we argue, then in a given
amount of integration time, the H$\alpha$ emission from an ionized halo should be detected (with
R=1000) with the same signal to noise as IR continuum imaging with
NGST for $z<6.6$. For $z>6.6$, due to the increase in dark current noise
in the $\lambda > 5 \mu$m regime, it will be detected with $10 \%$
of signal to noise ratio of IR imaging. Higher order lines such as
H$\beta$ should also be detectable. In addition, the SKA should be able to directly detect the same
ionizing sources in free-free emission. While the NGST will be able to
detect fainter sources than the SKA, the SKA has a larger field of
view: $\sim$200 times larger in solid angle. We find that the SKA
should be able to detect $\sim 10^{4}$ individual free-free emission sources with $z>5$ in
its field of view. However, a large population of
other radio bright sources (AGNs at low redshift, etc) will also be
detected. NGST (which can obtain redshift information from the
H$\alpha$ line)  must therefore be used in conjunction with the SKA to
distinguish between these two populations. 

Free-free sources below the flux limit of the SKA constitute a fluctuating background which can be detected statistically.
At arcsecond scales, Poisson fluctuations in the number of free-free sources
create temperature fluctuations detectable by SKA. At such small scales,
galactic foregrounds are expected to damp rapidly. We find that the
expected clustering of high-redshift ionizing sources makes a
negligible impact on the observed power spectrum of temperature
fluctuations. These temperature fluctuations have a highly non-Gaussian
signature. In particular, they should result in a detectable
skewness. Finally, ionized gas at high redshift may also be detected
by the kinetic Sunyaev-Zeldovich effect. While individual ionized halos are
probably too faint to be detectable by the SKA or MMA, the ionized IGM
around these halos create CMB anisotropies on sub-arcminute scales
which should be detectable by these instruments.     

I am grateful to my advisor, David Spergel, for his encouragement and
guidance. I also thank Michael Strauss for many detailed and helpful
comments on an earlier manuscript. Finally, I thank Nabila Aghanim,
Gillian Knapp, Alexandre Refregier and Stephen Thorsett for helpful
conversations. This work is supported by the NASA Theory program, grant number 120-6207.

\end{document}